\documentclass[preprint,12pt]{elsarticle}
\usepackage{amssymb}
\usepackage{amsmath}
\usepackage{color,array,amsthm}
\usepackage{graphicx}
\usepackage{comment}
\usepackage{hyperref}
\usepackage{xurl}
\usepackage[dvipsnames]{xcolor}
\usepackage{array,tabularx}
\usepackage{amsmath}
\usepackage{pifont}
\newcommand{\cmark}{\ding{51}}
\newcommand{\xmark}{\ding{55}}
\usepackage{multirow}
\usepackage{subcaption}
\usepackage{arydshln}
\usepackage{stackengine}
\usepackage{tikz}
\usepackage{cancel}
\usepackage{soul}
\usepackage{mathtools}
\usetikzlibrary{arrows.meta}
\definecolor{myorange}{RGB}{255,127,0}
\newcommand{\hmac}{\text{HMAC}}
\newcommand{\mac}{\text{MAC}}
\newcommand{\sig}{\text{sig}}

\makeatletter
\def\ps@pprintTitle{%
    \let\@oddhead\@empty
    \let\@evenhead\@empty
    \let\@oddfoot\@empty
    \let\@evenfoot\@empty}
\makeatother

\begin{document}

\begin{frontmatter}

%% Title, authors and addresses

%% use the tnoteref command within \title for footnotes;
%% use the tnotetext command for theassociated footnote;
%% use the fnref command within \author or \affiliation for footnotes;
%% use the fntext command for theassociated footnote;
%% use the corref command within \author for corresponding author footnotes;
%% use the cortext command for theassociated footnote;
%% use the ead command for the email address,
%% and the form \ead[url] for the home page:
%% \title{Title\tnoteref{label1}}
%% \tnotetext[label1]{}
%% \author{Name\corref{cor1}\fnref{label2}}
%% \ead{email address}
%% \ead[url]{home page}
%% \fntext[label2]{}
%% \cortext[cor1]{}
%% \affiliation{organization={},
%%             addressline={},
%%             city={},
%%             postcode={},
%%             state={},
%%             country={}}
%% \fntext[label3]{}

\title{CABBA: Compatible Authenticated Bandwidth-efficient Broadcast protocol for ADS-B}

%% use optional labels to link authors explicitly to addresses:
%% \author[label1,label2]{}
%% \affiliation[label1]{organization={},
%%             addressline={},
%%             city={},
%%             postcode={},
%%             state={},
%%             country={}}
%%
%% \affiliation[label2]{organization={},
%%             addressline={},
%%             city={},
%%             postcode={},
%%             state={},
%%             country={}}

\author[1]{Mikaëla Ngamboé}
\ead{mikaela-stephanie-2.ngamboe-mvogo@polymtl.ca}

\author[1]{Xiao Niu}
\ead{xiao.niu@polymtl.ca}

\author[2]{Benoit Joly}
\ead{Benoit.Joly@collins.com}

\author[2]{Steven P. Biegler}
\ead{Steven.Biegler@collins.com}

\author[4]{Paul Berthier}
\ead{p.berthier@rheagroup.com}

\author[3]{Rémi Benito}
\ead{Remi.Benito@aero.bombardier.com}

\author[2]{Greg Rice}
\ead{greg.rice@collins.com}

\author[1]{José M. Fernandez}
\ead{jose.fernandez@polymtl.ca}

\author[1]{Gabriela Nicolescu}
\ead{Gabriela.nicolescu@polymtl.ca}

%% Author affiliation
\affiliation[1]{organization={Department of computer and software engineering, Polytechnique Montréal Technological University},%Department and Organization
            city={Montréal}, 
            state={Québec},
            country={Canada}}

\affiliation[2]{organization={Collins Aerospace},%Department and Organization
            city={Cedar Rapids}, 
            state={Iowa},
            country={United States}}

\affiliation[3]{organization={Bombardier},%Department and Organization
            city={Montréal}, 
            state={Québec},
            country={Canada}}
            
\affiliation[4]{organization={Rhea Group},%Department and Organization
            city={Montréal}, 
            state={Québec},
            country={Canada}}
%% Abstract
\begin{abstract}
%% Text of abstract
The Automatic Dependent Surveillance-Broadcast (ADS-B) is a surveillance technology mandated in many airspaces.  It improves safety, increases efficiency and reduces air traffic congestion by broadcasting aircraft navigation data.  Yet, ADS-B is vulnerable to spoofing attacks as it lacks mechanisms to ensure the integrity and authenticity of the data being supplied.  None of the existing cryptographic solutions fully meet the backward compatibility and bandwidth preservation requirements of the standard.  Hence, we propose the Compatible Authenticated Bandwidth-efficient Broadcast protocol for ADS-B (CABBA), an improved approach that integrates TESLA, phase-overlay modulation techniques and certificate-based PKI. As a result, entity authentication, data origin authentication, and data integrity are the security services that CABBA offers.  To assess compliance with the standard, we designed an SDR-based implementation of CABBA and performed backward compatibility tests on commercial and general aviation (GA) ADS-B in receivers. Besides, we calculated the 1090ES band's activity factor and analyzed the channel occupancy rate according to ITU-R SM.2256-1 recommendation.  Also, we performed a bit error rate analysis of CABBA messages. The results suggest that CABBA is backward compatible, does not incur significant communication overhead, and has an error rate that is acceptable for Eb/No values above 14 dB.
\end{abstract}

%%Graphical abstract
%\begin{graphicalabstract}
%\includegraphics{grabs}
%\end{graphicalabstract}

%%Research highlights
%\begin{highlights}
%\item Implementing TESLA authentication into the ADS-B application layer and phase overlay modulation in its physical layer delivers a secure, backward-compatible, and bandwidth-efficient technology that meets the minimum operational performance standard for ADS-B
%\item Channel occupancy analysis shows that ADS-B authentication using TESLA adds minimal communication overhead, less than 6\%, when implementing phase overlay modulation in the ADS-B physical layer
%\item Backward-compatibility tests confirm phase overlay modulation in the ADS-B physical layer does not disrupt existing receivers, enabling seamless integration of new secure equipment without compromising ADS-B system safety
%\item Given the hardware and software architecture of most modern avionics systems, updating legacy ADS-B equipment to support TESLA authentication and perform phase overlay modulation could probably be done with a firmware and software upgrade.
%\end{highlights}

%% Keywords
\begin{keyword}
%% keywords here, in the form: keyword \sep keyword

%% PACS codes here, in the form: \PACS code \sep code

%% MSC codes here, in the form: \MSC code \sep code
%% or \MSC[2008] code \sep code (2000 is the default)
ADS-B, security, authentication, backward compatibility, bandwidth efficiency, TESLA, PKI.
\end{keyword}
\end{frontmatter}
\section{Introduction}

% WHAT IT ADS-B
Automatic Dependent Surveillance-Broadcast (ADS-B) is an aircraft
surveillance technology \cite{yang2022aircraft} that allows aircraft to broadcast information
about their identification, position, speed, and other data acquired
from onboard sensors \cite{adsbRTCA2012,RTCA260C2020,4444PANS}. 
It supports many airborne and ground safety applications \cite{CNSDef}.
For example, Air Traffic Control (ATC) can use ADS-B information as an
alternate means of surveillance, complementary to radar, to improve
efficiency of controlled airspace \cite{ED129,ED129A}. Furthermore, ADS-B
provides an alternate source of information to allow airborne aircraft
to maintain traffic separation.

% ADS-B SECURITY PROBLEM

% The ADS-B technology was originally designed in the early 1990s, to allow an aircraft to autonomously broadcast information about its position, speed and heading

ADS-B was originally designed in the early 2000s to replace radars as part of
the United States Federal Aviation Administration (FAA) NextGen
initiative \cite{NextGen}. It has since been adopted worldwide. 
Indeed, there are three certified ADS-B data links: the Universal Access Transceiver (UAT), which operates only in the United States at the 978~MHz frequency and uses 420-bit messages (272 bits for the payload); the 1~090~MHz Extended Squitter (1090ES), an internationally adopted link with 112-bit messages (56 bits for the payload); and the VHF Digital Link (VDL) MODE 4, which operates in the 108-136~975 MHz range with a message structure similar to that of the 1090ES, is most adopted in Northern Europe but is also rarely used due to the requirement for equipment upgrades \cite{datalink4ADS-B}.

Unfortunately, the ADS-B was conceived without any
communication security mechanisms \cite{costin2012ghost,manesh2017analysis}, which represents a significant threat to
aviation safety.    
Indeed, by using low-cost equipment such as a Software Defined Radio (SDR),
attackers could easily transmit false ADS-B messages
\cite{costin2012ghost, manesh2017analysis, strohmeier2014realities}  to create a confused and false picture of traffic for controllers and pilots. This can potentially lead to 
flight delays, separation conflicts between aircraft or unnecessary maneuvers by pilots.  In addition, spoofed ADS-B messages received and processed by Traffic Collision and Avoidance Systems
(TCAS) in the cockpit could affect the decision-making ability of air crews
\cite{ryon2018safety}.  Therefore, the use of ADS-B in ATC and traffic avoidance
can represent a security risk.

% ADS-B SECURITY REQUIREMENTS
Based on the foregoing, it is necessary to secure ADS-B.  In particular, to
prevent spoofing attacks, there must be a method to ensure \emph{identity
authentication} of the senders and \emph{message authentication} of transmitted
ADS-B messages.  This is achieved through the simultaneous fulfillment of these
three security goals:
\begin{enumerate}
    \item \emph{Data integrity}, is the assurance that data has not been altered in an unauthorized manner.
    \item \emph{Entity authentication}, also known as \emph{Identity Authentication},
      % AKA as Authentication or Identity Authentication
     is the assurance of the identity of a given entity interacting with a system. 
    \item \emph{Data origin authentication}, also known as
      \emph{Message Authentication}, is the assurance that a given
      entity was the original source of received data.
\end{enumerate}
% SECURE ADS-B OPERATIONAL REQUIREMENTS
Furthermore, any solution to secure ADS-B must adhere to the operational requirements delineated in the Minimum Operational Performance Standard (MOPS) for the 1~090~MHz frequency \cite{RTCA260C2020}, which serves as the primary channel for ADS-B communications. Specifically, the solution must be backward compatible with current receivers, ensuring their ability to accurately receive, interpret, and display position information for nearby traffic. Additionally, the solution is mandated to minimize the utilization of the congested 1090ES, which is extensively utilized by Secondary Surveillance Radar (SSR) and Extended Squitter (Mode S) transmitters, such as radar, multi-lateration, and airborne TCAS \cite{ED129, ED129A}.

% SOLUTIONS PROPOSED
Several cryptographic solutions have been proposed to secure ADS-B.  None of
them appear to meet all the security goals and operational requirements listed
above.  Therefore, we consider that the question of how to secure ADS-B while meeting security and operational constraints is still open.  

To that effect, in this paper, we introduce a solution called
Compatible Authenticated Bandwidth-efficient Broadcast protocol for
ADS-B (CABBA). CABBA integrates the Time Efficient Stream Loss-tolerant Authentication (TESLA) mechanism \cite{perrig2002tesla, perrig2003tesla} with phase overlay modulation techniques and a Public Key Infrastructure (PKI). By leveraging TESLA and PKI, CABBA fulfills all the security objectives specified earlier to prevent  ADS-B spoofing attacks. This includes data integrity, data origin authentication, and identity authentication. Furthermore, by integrating the phase overlay modulation in CABBA's physical layer, we aim to align our solution with the operational requirements outlined by the MOPS. %This includes backward compatibility with current ADS-B receivers and bandwidth preservation of the 1 090 MHz frequency.

% TRANSITION PERIOD AND BC REQUIREMENTS

Given the consequences of a potential attack exploiting the ADS-B vulnerability, one would hope that ADS-B be replaced as quickly as possible by a secure alternative. Unfortunately, such a one-to-one replacement will be lengthy and difficult in the context of aviation. First, it will likely take several years for an accepted standard to be drawn, discussed, approved, and then made mandatory by civil aviation authorities — at least 5 to 10 years. Second, considering the long lifetime of aircraft and their avionics, it is very likely that CABBA-capable and ADS-B legacy avionics would have to co-exist and use the same communication channels during the long transition period from initial deployment to full worldwide adoption. While it is paramount that CABBA-capable receivers be able to authenticate messages from CABBA-capable transmitters, it is equally important in terms of aviation safety that in the transition period both CABBA-capable and legacy ADS-B receivers be able to receive and interpret ADS-B messages from legacy ADS-B transmitters.

% RESEARCH QUESTIONS
In light of these operational requirements, the two most important questions regarding
any secure ADS-B solution, in particular CABBA, that needs to be answered are:
\begin{enumerate}
\item Could CABBA be gradually deployed while ensuring that legacy ADS-B equipment continues to operate?
\item What would be the viability of deploying CABBA in terms of communication
  channel saturation?
\end{enumerate}

To evaluate the backward compatibility of CABBA, we have constructed an SDR-based
implementation.  We have used this implementation to test backward compatibility
with two different suites of commercial-off-the-shelf (COTS) ADS-B In solutions:
One is used in General Aviation (GA), and the other is used in business jets and airline
transport aircraft.  We also used this lab implementation of CABBA to test and
analyze its bit error rate (BER).  A channel occupancy rate (COR) analysis was
also undertaken to quantify the channel occupancy overhead of CABBA in a likely real-world scenario. Besides, a safety impact assessment of unauthenticated messages was conducted to evaluate the effect of CABBA on the situational awareness of pilots and air traffic controllers.

%\textcolor{red}{
Considering the above discussion, the contributions of this work can be  summarized as follows:
\begin{enumerate}
    \item We introduce CABBA, a secure variant of ADS-B technology
      that is bandwidth-efficient, backward compatible, and offers an
      adequate level of security by providing simultaneously two
      security services: aircraft identity authentication and ADS-B
      message authentication.
    \item We use the D8PSK phase overlaid modulation technique, as
      defined in the MOPS, to support the transmission of additional
      security information required by CABBA while preserving
      bandwidth usage.
      To the best of our knowledge, this is the first proposal to use
      the phase overlay technique as specified in the MOPS.
    \item We performed tests on a commercial aviation avionics suite
      and with a general aviation ADS-B in receiver to check whether
      our solution would be backward compatible with legacy equipment.
    \item We carried out a channel occupancy analysis to verify the
      operational viability of our solution, in terms of channel
      occupancy.
    \item We conducted a safety impact of unauthenticated messages to assess their effects on the situational awareness of pilots and controllers.
    \item We provide a detailed specification of the CABBA protocol,
      including the structure of the different packet types (in-phase
      and quadrature), the authentication mechanism and the decision
      logic used to discriminate between genuine and false packets.
      This specification is sufficiently detailed to allow anyone to
      implement the CABBA solution, and serve as the basis for
      subsequent standardization and adoption by the aviation
      industry.
\end{enumerate}
%}

The remainder of the paper is structured as follows.  Section~\ref{section:RL}
reviews prior works on cryptographic approaches for securing ADS-B. Section~\ref{Section:Background} outlines the operational details of the TESLA protocol.
Section~\ref{section:phaseOverlay} describes how phase overlay modulation
techniques can be applied to ADS-B to increase data throughput while keeping the
channel activity rate constant.  Section~\ref{section:architecture} introduces
CABBA, a cryptographic approach for securing ADS-B that integrates the TESLA
authentication protocol with phase overlay modulation techniques.
Section~\ref{section:experimentalP} details the experimental procedures used to
assess CABBA backward compatibility and  Section~\ref{section:experimentalP2} the methodology for evaluating bit error, channel occupancy, and uncertainty delays.
We conclude in Section~\ref{section:conclusion} with a summary of our findings,
describing their consequences in terms of possible real-world deployment of CABBA
and highlight necessary future work in this direction.

\begin{table*}[ht!]
\scriptsize
\setlength{\tabcolsep}{3.7pt} % Adjust column separation for a compact look
\renewcommand{\arraystretch}{1.1} % Adjust row height
\caption{An overview of cryptographic techniques for enhancing ADS-B security.
The approaches are categorized into three groups based on their use of symmetric, asymmetric, or hybrid cryptography.}
\label{tab:Review}
\centering
\begin{tabularx}{\linewidth}{|l|p{0.4\linewidth}|c|c|c|c|c|}
\hline
 & Cryptographic primitive & \multicolumn{3}{c|}{Security goals} & \multicolumn{2}{c|}{Operational performances} \\
 
\cline{3-7}
 &  & Origin & Integrity & Entity & Backward & Bandwidth \\
 &  & Auth  &  & Auth & compatibility & preservation \\
\hline

\multirow{4}{*}{\rotatebox[origin=c]{90}{Symmetric}} & Encryption &  &  & &  & \\
 & \cite{finke2013ads,finke2013enhancing,huang2014enabling,abgeyibor2014evaluation,Habibi2023} & \xmark & \xmark & \xmark & \xmark & \cmark \\
\cline{2-7}

 & MAC &  &  & &  & \\
 & \cite{samuelson2006enhanced,kacem2015integrity} & \xmark & \xmark & \xmark & \cmark & \xmark \\
\hline

\multicolumn{1}{|c|}{\multirow{6}{*}{\rotatebox[origin=c]{90}{Asymmetric}}}
 & Digital signature using certificate-based PKI &  &  & &  & \\
\multicolumn{1}{|l|}{} & \cite{costin2012ghost,feng2010data,buchholz2013dpp} & \cmark & \cmark & \cmark & \cmark & \xmark \\
\cline{2-7}

\multicolumn{1}{|l|}{} & Digital signature using Identity-based PKI &  &  & &  & \\
\multicolumn{1}{|l|}{} & \cite{baek2013authentication,yang2013efficient,yang2014ebaa,yang2015new,he2016efficient,thumbur2019efficient} & \xmark & \xmark & \xmark & \cmark & \xmark \\
\cline{2-7}

\multicolumn{1}{|l|}{} & Digital signature using certificateless PKI &  &  & &  & \\
\multicolumn{1}{|l|}{} & \cite{braeken2019holistic,wu2019ads,asari2021new,subramani2021efficient} & \cmark & \cmark & \xmark & \cmark & \xmark \\
\hline

\multicolumn{1}{|c|}{\multirow{6}{*}{\rotatebox[origin=c]{90}{Hybrid}}}
& Encryption using TESLA with Certificate-based PKI &  &  & &  & \\
\multicolumn{1}{|l|}{} & \cite{yang2017lhcsas} & \cmark & \cmark & \cmark & \xmark & \xmark \\
\cline{2-7}

\multicolumn{1}{|l|}{} & MAC or Digital signature using TESLA with certificate-based PKI &  &  & &  & \\
\multicolumn{1}{|l|}{} & \cite{berthier2017sat,sciancalepore2018sos} & \cmark & \cmark & \cmark & \cmark & \xmark \\
\cline{2-7}
\multicolumn{1}{|l|}{} & MAC using TESLA with certificate-based  &  &  & &  & \\
\multicolumn{1}{|l|}{} & PKI and phase overlay techniques &  \cmark & \cmark & \cmark & \cmark & \cmark \\
\hline
\end{tabularx}
\end{table*}
\section{Overview of cryptographic solutions for ADS-B} \label{section:RL}
In this section, we review previous works and characterize the security goals and operational performance requirements they did not meet. We group these works into three categories, they use symmetric, asymmetric, and hybrid cryptography. 

\subsection{Symmetric cryptography-based protocols} 

The studies that use symmetric cryptography to secure the ADS-B rely on cryptographic primitives such as encryption or message authentication code.

Format-preserving encryption, or FPE, involves encrypting data in a manner such that the resulting \emph{ciphertext} preserves the format of the original \emph{plaintext} \cite{NISSP80038GRev1}.
Some studies employ this approach because it aligns with the technological requirements of the ADS-B standard in preserving the bandwidth of the 1090ES channel \cite{finke2013ads, finke2013enhancing, huang2014enabling, abgeyibor2014evaluation, Habibi2023}. However, encryption schemes fall short of meeting the backward compatibility criteria of the ADS-B standard, primarily because navigation data are not transmitted in \emph{plaintext}. To overcome that limitation, it has been suggested to use instead message authentication codes or MAC \cite{samuelson2006enhanced,kacem2015integrity}. For the MAC approach to be effective, there must be symmetric trust assurance between the communicating parties. However, it is challenging to achieve in open communications such as that of the ADS-B because it is often impossible to manage and master the parties involved in the broadcast. In such a scenario, knowing that when employing symmetric cryptography every receiver must know the symmetric key,  a malicious actor can impersonate a sender and forge messages to other receivers.
 
 To ensure authenticated broadcast, ADS-B requires an asymmetric process enabling every receiver to ascertain the genuineness of received messages, devoid of the ability to produce genuine messages from received ones \cite{Challal2004MulticastAuth}. Asymmetric cryptography, particularly digital signature, is the standard technique to achieve this \cite{DigitalSignatureStandard}.
 
\subsection{Asymmetric cryptography-based protocols}

In \emph{asymmetric or public-key cryptography}, a pair of keys (public and private) is used for encryption and digital signatures. The \emph {private key} is kept secret, while the \emph{public key} is shared for secure communication. To guarantee authenticity, the public key must be confirmed by a certification authority (CA) through a public key certificate that links the key to an entity \cite{NISSP800175BRev1}.
The system responsible for issuing, maintaining, and revoking certificates is known as a PKI \cite{FIPS186-5,EncyclopediaCryptoandSec2011,NISTSP80057}. For ADS-B, the literature proposes three types of PKI to manage aircraft certificates: certificate-based, identity-based, and certificateless.

Among the certificate-based PKI solutions proposed to secure ADS-B is an authentication scheme that relies on Elliptic Curve Digital Signature Algorithm (ECDSA) signatures and X.509 certificates \cite{feng2010data}. Although this solution might fulfill the demands of the ADS-B protocol regarding security, it fails to meet the standard's technological performance criteria. In addition, the authors leave open the issue of certificate distribution and do not address that of certificate revocation. To address the weaknesses of \cite{feng2010data}, a lightweight PKI solution is recommended in \cite{costin2012ghost}, where the ADS-B message is signed, and its signature is partitioned across \emph{N} messages. It is suggested that keys distribution occurs during the routine maintenance of the aircraft. Furthermore, still to address the limitations of \cite{feng2010data}, a dual path PKI solution that aims to handle the certificate revocation problem by using session certificates is proposed in \cite{buchholz2013dpp}. According to this scheme, an aircraft should have certificates from both their home country's National Aviation Authority (NAA) and the local ATC center where they are currently located. Thus, the dual certification is evidence that the aircraft has been granted permission to fly, as well as validated as a safe and current entity within the local center from which it is flying. We argue that the adoption of the PKI proposed in \cite{buchholz2013dpp} will raise the operational expenses of the ADS-B system and render its use cumbersome, especially for international flights. In general terms, using certificate-based PKI for ADS-B security has two limitations. First, it significantly increases communication costs, conflicting with the bandwidth preservation criteria of the ADS-B standard, proof of which is that none of the above-mentioned solutions \cite{costin2012ghost,feng2010data,buchholz2013dpp} meet this need. Second, establishing and operating a PKI for these solutions is impractical in the current state of global coordination among ICAO and NAAs.

Identity (ID)-based PKI attempts to eliminate the key distribution problem of certificate-based PKI. In ID-based PKI, public keys are derived from easily identifiable user attributes, such as email addresses, eliminating the need for traditional certificates and the complex infrastructure that supports them \cite{shamir1984identity}. This is achieved through a central entity called a private key generator (PKG), tasked with computing each user's private key based on their corresponding public key \cite{shamir1984identity,hu2010cryptanalysis}.
Several studies have used the ID-based authentication approach to secure ADS-B communications. For instance, a scheme that signs ADS-B messages in two stages, online and offline, has been developed to increase the efficiency of the signature generation process \cite{baek2013authentication}. Furthermore, a broadcast authentication technique incorporating batch verification of digital signatures \footnote{Batch verification allows to simultaneously verify multiple digital
signatures, whether they were produced by one signer or several.} has been proposed to reduce the time and computational expense involved in the signature verification process for ADS-B messages \cite{yang2013efficient}. Subsequently, a broadcast authentication protocol that relies on ID-based signature and enables message recovery has been designed \cite{yang2014ebaa}. Aware that working with a single PKG in large-scale endeavors is not viable \cite{chow2004secureHIBS}, the authors of the contribution \cite{yang2015new} took inspiration from the hierarchical ID-based cryptosystems \footnote{In a hierarchical ID-based cryptosystem, multiple PKGs create a tree-like structure \cite{gentry2002hierarchical,chow2004secureHIBS}. The primary PKG generates private keys for its subordinates, who, in turn, produce private keys for PKGs beneath them \cite{gentry2002hierarchical,chow2004secureHIBS}. PKGs at the edges generate private keys for users  \cite{gentry2002hierarchical,chow2004secureHIBS}.} 
presented in \cite{gentry2002hierarchical,chow2004secureHIBS} and implemented an authentication framework that relies on hierarchical ID-based signature (HIBS) and performs signature batch verification. However, the need for intricate hash-to-point operations during signature and verification processes renders the scheme \cite{yang2015new} non-lightweight, reducing its deployability. To overcome this limitation, a three-level hierarchical ID-based signature scheme (TLHIBS) that relies solely on general hash functions has been introduced in \cite{he2016efficient}. Despite this effort, the issue of computational overhead persisted. In response, an alternative scheme that avoids employing any intricate bilinear pairing operations over elliptic curves has been implemented in \cite{thumbur2019efficient}. This approach slightly reduces the computational overhead when compared to the previous works \cite{baek2013authentication,yang2013efficient,yang2014ebaa,yang2015new,he2016efficient}. Besides, all these solutions have two additional drawbacks. First, they increase communication overhead, violating ADS-B bandwidth requirements for the 1090ES channel, which makes them unimplementable. Second, they are vulnerable to key escrow, a privacy issue in ID-based cryptosystems, allowing an untrustworthy PKG to decrypt messages and forge signatures by accessing users' secret keys \cite{asari2021new}. This vulnerability raises substantial concerns about the overall security of these proposed solutions.

Certificateless PKI eliminates the key escrow problem by splitting the private key generation process between the PKG and the user. The PKG generates a portion of the private key, while the user creates a random value for the remaining portion, which is kept confidential. This approach has been used to implement ADS-B messages authentication schemes that rely on certificateless short signatures \cite{braeken2019holistic,wu2019ads}. These schemes were subsequently enhanced by integrating privacy-preserving and aggregate signature methods to ensure sender anonymity and reduce the computational cost of signature verification \cite{asari2021new,subramani2021efficient}. The concept of certificateless short signature is new, and while it appears promising, it is not yet mature enough to be adopted. Indeed, a significant challenge in certificateless cryptography lies in the establishment of security schemes that can ensure a satisfactory level of protection against attackers attempting to manipulate users into employing counterfeit public keys. This difficulty arises from the absence of digital certificates to unequivocally verify the authenticity of a public key \cite{Dent2011}.

\subsection{Hybrid cryptography-based protocols}
So far, we have seen that there are two approaches to secure ADS-B while adhering to the standard's backward compatibility criteria. Through MACs using symmetric cryptography or through digital signatures using asymmetric cryptography, notably that based on certificate-based PKI. The digital signature approach is secure, however, the generated signatures are too long, which causes problems if we consider the requirement of preserving the 1090ES's bandwidth. On the other hand, the MAC approach allows generating short signatures, nevertheless, it is not secure since symmetric trust cannot be ensured between communicating parties.
As a result, some authors have proposed using hybrid cryptography, particularly the Timed Efficient Stream Loss-tolerant Authentication (TESLA) protocol  \cite{perrig2003tesla}. Details of how TESLA operates can be found in Section \ref{Section:Background}.

Security in the Air using TESLA or SAT, is an authentication protocol that adapts TESLA to the requirements of ADS-B~\cite{berthier2017sat}. It replaces TESLA's synchronization protocol with onboard GPS clock time and employs certificate-based PKI for aircraft and message authentication. SAT, tested on \texttt{gr-air-modes}, shows potential backward compatibility with existing ADS-B receivers. However, it has two limitations. Firstly, it increases bandwidth usage by requiring three types of packets for message authentication. For simplicity of explanation, we refer to them as Type A, B and C packets. Standard 112-bit ADS-B packets are replaced with Type A packets that include a 16-bit MAC code and 8-bit sequence number, increasing to 136 bits (a 14\% increase). Type B packets, containing TESLA authentication keys, are 184 bits long, and Type C packets containing aircraft certificates are 1520 bits long. Let $\Delta_B$ be the time between transmission of Type B packets (originally set to 5 seconds), and $\Delta_C$ be the time between transmission of certificate packets (originally set to 30 seconds). Assuming a mean transmission rate $\bar{f}_A$ of 6.2 ADS-B messages per second per aircraft, the use of SAT results in an additional transmission overhead per aircraft per minute given by:
\begin{equation}
\begin{split}
O_{\min} &= (\bar{f}_A \cdot 60 \cdot 24) + \left(\frac{60}{\Delta_B} \cdot 184\right) + \left(\frac{60}{\Delta_C} \cdot 1520\right) \\
&= 14752.
\end{split}
\end{equation}
This results in a total overhead of 245.8 bps over the normal bit rate of 694.4 bps for standard ADS-B message transmission, a total 35\% increase in bandwidth usage.
The second limitation is related to security.  To limit bandwidth usage, the authors of SAT limited the size of the MAC to 16 bits.  Truncating the MAC like this is a standard described and accepted by FIPS standard 198-1 \cite{HMACStandard} and described in FIPS Standard Publication 800-107 \cite{HashStandard}.  In this case, the residual attack risk is two-fold:
\begin{enumerate}
\item The attacker is lucky and guesses the right MAC for a spoofed message he desires to send.  This will happen with probability $2^{-16}$.
\item The attacker floods the channel with spoofed messages with all MAC possibilities, i.e. he sends $65 536=2^{16}$ messages hoping that the ADS-B receivers ignore the ones with a wrong MAC and process and accept the one with the correct.
\end{enumerate}
In an ideal scenario where bandwidth is not constrained, we believe a larger MAC size would provide better security, ideally with a minimum of 32 bits, forcing the attacker to be extremely lucky or have to send an astronomical number of messages (over 4 billion messages) for his attack to be successful.  

Securing Open Skies or SOS, is a solution that integrates TESLA with a mechanism for collectively verifying all messages transmitted by an aircraft within a specified timeframe \cite{sciancalepore2018sos}. Unlike the SAT method, which authenticates messages individually using MAC, SOS opts for batch authentication through digital signatures. This strategy is designed to effectively tackle the bandwidth consumption limitations of SAT and the broader challenge of bandwidth constraints in the 1090ES band.
However, although transmitting one digest per message pool takes less bandwidth, the SOS technique can be troublesome in some instances. In the case of message injection, for example, the receivers must get the set of genuine messages. The authors propose a community server-based majority voting filtering stage. To identify the correct message sequence, servers try various message combinations as well as hash operations and comparisons. We argue that if an attacker injects false messages at a high rate, it will result in computation and a time-consuming task. Furthermore, should any of the ground receivers fail to receive a single packet, all packets delivered during that interval cannot be validated, posing a serious safety issue. 

The solution presented in \cite{yang2017lhcsas}, combines Format-preserving, Feistel-based encryption, and TESLA to ensure the confidentiality and integrity of ADS-B messages. However, due to the lengthy security parameters required in their authentication technique, their solution necessitates the transmission of five ADS-B messages for every navigation data sent by an aircraft. This results in significant bandwidth consumption, thereby failing to meet the bandwidth preservation requirement outlined in the standard.
Additionally, their proposed encryption of the ICAO code contradicts backward compatibility criteria. Consequently, this solution fails to comply with any of the operational requirements specified in the MOPS for ADS-B.

\section{Background} \label{Section:Background}

In this section, we give a detailed explanation of how TESLA \cite{perrig2003tesla} protocol operates.

\subsection{Timed Efficient Stream Loss-tolerant Authentication (TESLA)} \label{subsection:TESLA}
The TESLA protocol combines asymmetric and symmetric cryptography to capitalize on their respective advantages.  The core concept underlying TESLA is that the sender, Alice, adds to every packet a MAC computed with a secret authentication key $K'$ known only by her. The receiver, Bob, buffers the packet when it arrives because he lacks the key to authenticate it.  Only when Alice sends it to him, a while later, will he be able to verify the authenticity of the packet. To function properly, TESLA requires time synchronization of senders and receivers and, a trustworthy method for producing keys at the sender and authenticating them at the receiver.  

The authentication keys are generated by the sender, Alice, before the broadcast begins.  First, she divides the broadcast period into $N$ time intervals.  Second, she constructs a one-way keychain of length $N$, the last key generated $K_0$, serves as a pledge spanning the whole chain and may be used to verify any of the keys in the chain using the formula $K_0=F^i(K_i)$. Third, she applies a one-way function $F'$ to the keys of the keychain.  This process generates the TESLA \emph{authentication keys} $K_i'=F'(K_i)$, which are used to calculate the MAC of the messages to be broadcast.

Before starting to broadcast,  Alice communicates the key disclosure delay $d$ and the pledge to the keychain $K_0$ to Bob, the receiver, via a secure channel. Then, to broadcast a message $m_j$ at time interval $i$, Alice must first compute the $\mac=\mac(K_i', m_j)$, then build the TESLA packet $P_j$ which is then broadcast.
\begin{equation}
  P_j= m_j \, \| \, \mac(K_i', m_j) \, \| \, K_{i-d}
\end{equation}

When Bob receives $P_j$, he stores the triplet ($i$, $m_j$, $\mac(K_i', m_j)$) in a buffer while waiting for the TESLA interval key that will allow him to deduce the authentication key $K_{i}'$ and validate the MAC of the message $m_j$.  Furthermore, Bob checks the authenticity of the origin of the interval key $K_{i-d}$ by determining whether there exists a small integer $v$ (i.e. of size commensurate with the number of intervals in a typical flight) such that $K_0=F^v(K_{i-d})$ .  In such an event, Bob computes the authentication key $K_{i-d}'= F'(K_{i-d})$ and then validates the integrity of the messages broadcast within the time interval $i-d$ by computing their MAC and comparing them with the stored ones.
\section{Phase overlaid modulation techniques} 

\label{section:phaseOverlay}

\begin{figure*}[ht] 
\includegraphics[width=1.0\textwidth]{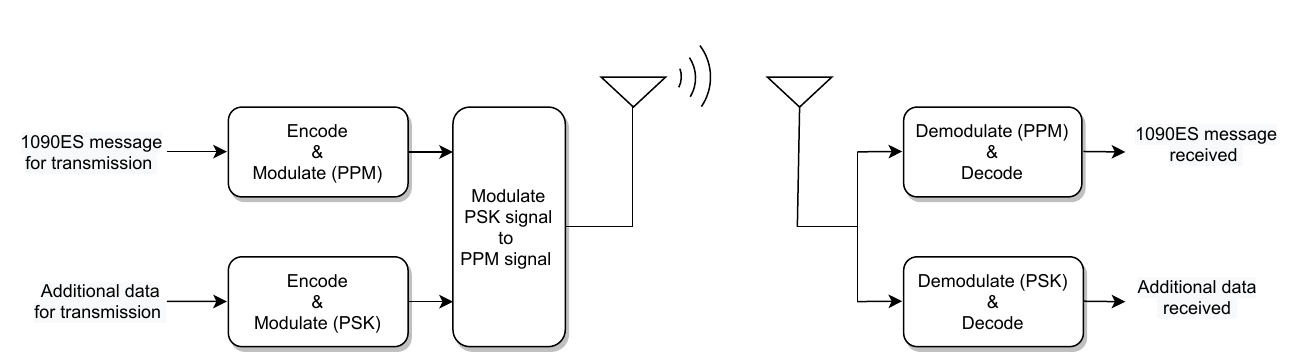}
\caption{Block diagram for the use of phase overlay method to add more data to a 1090ES message, as explained in the patents \cite{EP2661039B1, US20100079329A1}. To perform an SDR-based implementation using an I-Q modulator at the input of the transmitting antenna and an I-Q demodulator at the output of the receiving antenna is the most practical way to proceed. In this way, the 1090ES message is conveyed in the in-phase (I) component of the carrier and the additional data in its quadrature (Q) component.}
\label{fig:phaseOverlaidFig}
\end{figure*}

In this section, we focus on phase overlay modulation techniques and describe how they can be applied at the physical layer of ADS-B to increase data throughput while keeping the channel activity rate constant.

Systems that are currently envisioned by avionics system designers will most likely require more data transmission than the 6.2 messages per second restriction allowed by the ADS-B standard \cite{EP2661039B1, RTCA260C2020}. Furthermore, increasing data throughput is a \emph{sine qua non} condition for securing the protocol. Both industry and academia are aware of this need and have begun to look for methods to increase data throughput while meeting the standard's requirements of preserving the 1090ES band \cite{RTCA260C2020, Yeste-pskModulation-2015, EP2661039B1, Zambrano-PSK-2018,Leonardi-phaseMod-2020}.  There are three versions of ADS-B, with the most recent (Version 3) released in the 2020 MOPS \cite{RTCA260C2020}. This version of the MOPS incorporates the notion of phase overlay capacity, which involves using alternate modulation techniques to increase data throughput without increasing channel activity rate. Although phase overlay is not required in this version of the MOPS, it is included so that stakeholders can begin designing, manufacturing and testing equipments and systems with the capability\cite{Arbuckle2021}.

The MOPS proposes the use of the phase overlay functionality to encode additional bits of information into a conventional 1090ES message beyond the original 112 bits.  The phase overlay method proposed is that described in a patent \cite{EP2661039B1}.  As depicted in Figure~\ref{fig:phaseOverlaidFig}, this can be done by performing a pulse position modulation (PPM) on the 1090ES message to be transmitted, then performing a phase shift keying (PSK) modulation on the additional data to be transmitted. To complete the process, the PSK signal resulting from the previous step has to be modulated to the PPM signal resulting from the first step.

PSK is a modulation technique in which data is transmitted by altering the phase of the carrier wave. It was chosen as the overlay modulation method because it can be individually demodulated (Figure \ref{fig:phaseOverlaidFig}) and is non-destructive to the original message sent by amplitude modulation \cite{EP2661039B1,US20100079329A1}. In principle, changing the phase of the carrier signal should not affect the older hardware's ability to decode the original 1090ES message \cite{EP2661039B1,US20100079329A1}.  However, although there is agreement on the usage of PSK modulation as the overlay modulation technique, stakeholders are still divided on which combination to adopt. The study \cite{Leonardi-phaseMod-2020} suggest using the binary phase shift keying (BPSK) method, which allows doubling the throughout, sending a total of 224 (112*2) bits. However, this amount of bits is insufficient when we consider that the smallest digests produced by SHA-2 (SHA-224) and SHA-3 (SHAKE 128) are 224 bits and 128 bits, respectively \cite{SHA2,SHA3}.  Furthermore, the smallest ECDSA
signature is 256 bits \cite{DigitalSignatureStandard}. It is precisely in order to allow signing of ADS-B messages that researchers \cite{Yeste-pskModulation-2015, Zambrano-PSK-2018} have advocated using the Differential 16-Phase-Shift Keying (D16PSK) as it allows quintupling the throughput. However, as is widely known, increasing the modulation order increases BER.  This tradeoff in transmission reliability is probably one of the reasons why the RTCA \cite{RTCA260C2020} advocates using D8PSK modulation, in combination with error correction codes such as Reed-Solomon or Low-Density Parity-Check.  As a result, only 204 of the 336 extra bits provided by D8PSK can be used to convey extra information.  Thus, while phase overlay techniques increase data capacity, there are still certain constraints when it comes to securing ADS-B communications.  Traditional digital signatures using certificate-based PKI continue to be a concern regarding the communication cost, and this despite the increase in data capacity. In contrast, short signatures employing hybrid cryptography appear to be a more attractive option.
\section{CABBA : Compatible Authenticated Bandwidth-efficient Broadcast protocol for ADS-B } \label{section:architecture}

The CABBA solution is presented in this section. CABBA seamlessly integrates phase-overlay modulation into the ADS-B physical layer and TESLA authentication into its application layer, while using a certificate-based public key infrastructure.  CABBA is based on a new approach in which the fundamental TESLA concept of MAC-based message authentication and key disclosure delay remains intact, while introducing a significant transformation in the way information is transmitted and the type of information transmitted to enhance security. In the following lines, we delve into how CABBA distinguishes itself from previously proposed TESLA-based solutions, i.e.~1) the integration of phase overlay modulation and 2) enhanced packet structure.

\paragraph{Integration of Phase Overlay Modulation}
First, we enhance the physical layer of ADS-B by incorporating the phase overlay modulation technique proposed in the patent \cite{EP2661039B1} recently promoted by the RTCA in the most recent version of the ADS-B MOPS \cite{RTCA260C2020}.  As detailed in Figure~\ref{fig:phaseOverlaidFig}, part of the information is conveyed in the in-phase component of the carrier, and the remaining information is in its quadrature component.  For this purpose, we consider the two phase overlay modulation techniques mentioned earlier:
\begin{enumerate}
  \item the D8PSK method advocated by RTCA \cite{RTCA260C2020}, and
  \item the D16PSK method proposed by academia \cite{Yeste-pskModulation-2015, Zambrano-PSK-2018}.
\end{enumerate}

To determine which of these techniques is most appropriate for CABBA, we
first implemented and conducted backward compatibility tests with both of them.
Most importantly, we set to identify which of these modulation techniques
provides an optimum trade-off between higher data throughput and acceptable
quality of signal, i.e.~a lower BER, by conducting a simulation study.  This is
described in Section~\ref{section:experimentalP}.  Nonetheless, since D8PSK is
the proposed standard and for the sake of simplicity of explanation, in the rest
of this section we describe only the implementation with D8PSK.  This
configuration allows for an additional 336 (3*112) bits to be sent together with
the ADS-B 112-bit original message.

\paragraph{Enhanced Packet Structure}
\begin{table*}[ht]
\hskip-1.4cm
\scriptsize 
 \begin{tabular}{|l|l |l |l | l | l | c | c | c| c |}
    \hline
     \multicolumn{4}{|c|}{Type}  & Content & Period & \multicolumn{4}{c|}{Size (bits)} \\\cline{1-4}\cline{7-10}
     \multicolumn{1}{|c|}{TESLA} & \multicolumn{1}{c|}{SAT} & \multicolumn{1}{c|}{SOS} & \multicolumn{1}{c|}{CABBA} & & & \multicolumn{1}{|c|}{TESLA} & \multicolumn{1}{c|}{SAT} & \multicolumn{1}{c|}{SOS} & \multicolumn{1}{c|}{CABBA} \\\hline
     A &   &    &   &Message, MAC &  - & - & - & - & - \\ \hdashline
      &  A &    &   &ADS-B message, MAC, sequence no. &  - & - & 136& - & - \\ \hdashline
      &    & A1 &   &ADS-B message &  - & - & - &112 & - \\ 
      &    & A2 &   &MAC & $T_{A2}$ & - & - &128 & - \\ \hdashline
      &    &    & A &ADS-B message, MAC, sequence no. &  - & - & - & - & 112 \\ \hline
    B &    &    &   & Interval key & $T_{B}$  & 128 & - & - & - \\ \hdashline
      &  B &    &   & Interval key & $T_{B}$  & - & 184 & - & - \\ \hdashline
      &    & B  &   & Interval key & $T_{B}$  & - & - & 128 & - \\ \hdashline
      &    &  & B1  & Interval key & $T_{B1}$ & -   & -& -& 128\\ 
      &    &  & B2  & Interval key and signature & $T_{B2}$ & -   & -& - & 210 \\ \hline
      &  C &    &   & \parbox[t]{5cm}{Interval key and signature; \\ aircraft public key and signature} &$T_C$ & - & 1520 & - &-\\ \hdashline
      &    &    &  C &aircraft public key and signature &$T_C$ & - & - & - &242\\ \hline
    \end{tabular}
    
 \caption{A Comparison of CABBA's packet structure with that of earlier TESLA-based solutions. In CABBA, Type B packets are replaced with Type B1 Packets at the beginning of each interval (each $T_B=T_{B1}$ seconds) and by Type B2 packets every $T_{B2}$ seconds.  Type C packets are shorter than in SAT and sent with period $T_C$.}
 \label{tab:CABBApacketTypes}
\end{table*}

The second distinction between CABBA and previous works relates to the content and structure of the packets to be transmitted, as highlighted in Table \ref{tab:CABBApacketTypes}. 

In TESLA, the security information in Type A packets (the MAC) and the interval key subsequently received via Type B packets allows the receiver to verify \emph{data integrity} of the message. \emph{Data origin authentication} is achieved by cross-referencing the information in Type B packets with additional data shared from sender to receiver via a secure communication channel. Both protocols, TESLA \cite{perrig2003tesla} and SOS \cite{sciancalepore2019sos}, assume pre-existing trust between communicating parties, presupposing that the receiver possesses the sender's certificate beforehand. 

In real-life operation, however, an aircraft cannot anticipate precisely which other planes it will encounter along its flight path.  Consequently, the authors of SAT \cite{berthier2017sat} propose a practical solution: distributing certificates through type C packets—an approach we endorse. Nonetheless, there are at least three aviation scenarios in which Bob (the receiver) may not require the certificate, as he may already possess the sender's public key and have duly authenticated its legitimacy. These scenarios are:

\begin{enumerate}
    \item Bob, as an ADS-B ground station, receives messages from aircraft either directly (via Line-of-Sight RF signal) or indirectly (via satellite). To authenticate these messages, Bob's ground station can access a PKI containing public keys of worldwide aircraft, indexed by their ICAO ID. This setup enables instant entity authentication without relying on aircraft to transmit certificates.
    \item Alice, another ADS-B ground station, transmits information to airborne aircraft like Bob through LOS signal. Here, it's reasonable for the aircraft's receiver to hold a small, infrequently updated database of public-key certificates provided by the NAA (e.g., FAA) for authentication.
    \item Air-to-air ADS-B transmissions pose challenges as it's impractical to preload worldwide aircraft public keys into each aircraft's receiver, let alone update them frequently due to aircraft turnover. However, future adoption of integrated digital communications like ATN could enable real-time access to remote PKIs, allowing aircraft to cache recently encountered aircraft's public keys for authentication.
\end{enumerate}

To accommodate for such situations where the transmission of certificates might not be needed or not be needed as often, we propose the packet structure that follows :

\begin{description}
\item[\textbf{Type A.}] Contain the ADS-B message its MAC and sequence number under the interval
\item[\textbf{Type B1.}] Contain the interval key $K_i$ 
\item[\textbf{Type B2.}] Contain $K_i$ and the digital signature of $K_i$.
\item[\textbf{Type C.}] Contain the aircraft public key $K_{pub}$ and its signature by the CA
\end{description}

Unlike SAT \cite{berthier2017sat}, Type C packets in CABBA do not contain interval key information or interval key signatures, and are therefore shorter. Signed interval keys are transmitted in a new type of packet, Type B2, which contains only an interval key and its signature. This has the advantage that signed keys can be sent with a lower frequency than certificates, resulting in a better use of bandwidth. In addition, this CABBA packet structure has the advantage of reducing bandwidth usage by eliminating the redundant transmissions of the interval keys, as it was the case in SAT \cite{berthier2017sat}.

\subsection{CABBA on the sender side}
CABBA requires airplanes to have a private-public key pair ($K_\text{pr}$, $K_\text{pub}$) and a certificate issued by a well-known and trusted certification authority.  Before the flight begins, Alice, the sender, divides its duration into equal intervals of $d$ seconds, and generates an authentication key for each interval.  The process for generating these keys is the same as that used in TESLA and SAT.  During the flight, the ADS-B messages and their MAC, the authentication keys of the intervals, and the certificate of Alice's aircraft are sent as described below.  

\subsubsection{Sending a message and its MAC}
To send an ADS-B message $m$ at time interval~$i$, Alice first produces the security data $\sigma$ for message $m$. This includes:
\begin{enumerate}
\item The message MAC, formed by the $\lambda$ leftmost bits of the message $\hmac(m,K_i')$
 \item The message sequence number $s$ for $m$ within that interval $i$
\end{enumerate}

In other words $\sigma=\mac \, \| \, s$.
%
%Let $m'$ represent the 120-bit packet containing the 112-bit ADS-B message payload $m$ and the fixed 8-bit preamble.  
As before, this information will continue to be encoded into the in-phase component of the RF signal.  We denote by $P_{A-I}=m$ the message information sent in-phase.  The security information $\sigma$ will be sent using the quadrature component of the RF signal and is thus denoted $P_{A-Q}=\sigma$.

With the same packet length (112 bits) and containing the same information encoded in the same manner as standard ADS-B packets, $P_{A-I}$ packets are intended to be fully intelligible by legacy ADS-B receivers.  A logical packet $P_{A-Q}$, on the other hand, will in principle only be intelligible with CABBA-compliant receivers.  With the choice of D8PSK, the highest quantity of bits that can be encoded in the quadrature component is 336 bits.  Nonetheless, not all of these bits are available to encode the security information $\sigma$.
The RTCA recommends using 12 bits to encode a reference phase and 120 parity bits to support the \texttt{RS(54,34)} error-correcting code, which must be applied to the $\sigma$ security data.  This means a maximum size of 204 bits for $\sigma$, which with the 8-bit sequence number $s$ results in a maximum size of 196 bits for the MAC, i.e.~$\lambda\leq 196$.

Alice uses the logical packet $P_{A-I}$ to perform PPM on a pulse train to generate the signal $S_{A-I}(t)$ as follows:
\begin{eqnarray} \label{eq:PPMmod}
  S_{A-I}(t) & = &\sum_{k=0}^{111}g(t-t_k) \quad ; \nonumber \\
  t_k & = & kT_S + m_{t}(1-P_{A-I}(k))
\end{eqnarray}
where $T_S=1\,\mu s$ is the symbol period for 1090ES transmissions,
$m_t=T_S/2$ is the PPM time-modulation index and $P_{A-I}(k)$ is
the value of the $k$-th bit which will be transmitted at time $k*T_S$.
Alice simultaneously uses logical packet $P_{A-Q}$ to perform D8PSK modulation on a sine wave to generate the signal $S_{A-Q}(t)$ as follows:
\begin{eqnarray}
  S_{A-Q}(t) &=& \sum_{k=0}^{111}\sin{(\omega_{c}t-\theta_k)} \quad ; \nonumber \\
  \theta_k &=& \frac{2\pi}{8} \; \text{symbol}_{P_{A-Q}}(k)  \label{eq:DPSKmod}
\end{eqnarray}
where $\omega_c$ represents the carrier signal frequency, $\theta_k$ is the phase associated to the 8PSK symbol $\text{symbol}_{P_{A-Q}}(k) \in [0,7]$, which is computed from the three bits from $P_{A-Q}$ to be transmitted at time $k*T_S$. 

\begin{equation}
\begin{split}
\text{symbol}_{P_{A-Q}}(k) = & 2^2P_{A-Q}(3k) + 2^1P_{A-Q}(3k+1) +\\
& 2^0P_{A-Q}(3k+2)
\end{split}
\end{equation}

These two signals $S_{A-I}$(t) and $S_{A-Q}$(t) are then used by Alice
to I-Q modulate (Equation~\ref{eq:IQmod}) the 1090ES carrier and so,
produce the radio signal $S_A$ to be broadcast.
\begin{equation} \label{eq:IQmod}
    S_A(t)=S_{A-I}(t)\cos(\omega_{c}t) + S_{A-Q}(t)\sin(\omega_{c}t)
\end{equation}

\begin{figure*}[t!] 
\centering
\includegraphics[width=0.9\textwidth]{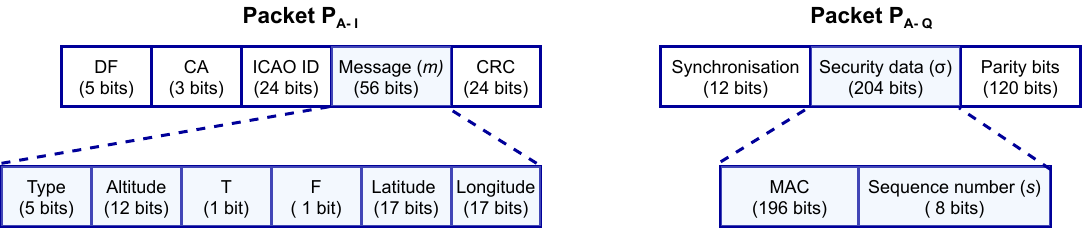}
\caption[Structure of the Type A packets in CABBA.]{Structure of the Type A packets in CABBA.  The ADS-B message $m$ encoded in the in-phase component $P_{A-I}$ (in this example an airborne position report) and the security data $\sigma$ in the quadrature component $P_{A-Q}$.  These two logical packets are then used to generate the in-phase and quadrature signal components $S_{A-I}$ and $S_{A-Q}$ of the RF signal to be transmitted $S_{A}$.}\label{fig:PA}
\end{figure*}
  
\subsubsection{Sending authentication keys and their signatures}
In order to allow the receiver to authenticate the Type A messages sent in interval $i$, the sender must later disclose the corresponding interval keys and their signatures.  This is done by sending Type B1 and B2 packets in subsequent intervals.  

Type B1 packets contain the TESLA interval $K_i$ (128 bits) from which the authentication key $K_i'=F'(K_i)$ of the interval $i$ is calculated.  The corresponding packet $P_{B1}$ will be transmitted during the next time interval $i+1$.  These packets are sent at the beginning of each interval, i.e.~every $T_{B1}=T_{int}$ seconds.

The signature of the authentication keys is added in Type B2 packets.  B2 packets replace B1 packets at the beginning of the interval, every fixed number $k$ of intervals.  Their transmission period $T_{B2}$ is thus a multiple of $T_{B1}$, with $T_{B2}=kT_{B1}$.  A typical packet $P_{B2}$ of this type will contain:
\begin{equation}
P_{B2} = K_i \, \| \, \sig_{K_{pr}}(K_i)
\end{equation}
where the $\sig_{K_{pr}}$ represents the chosen signature-generating function with private key $K_{pr}$.  

\begin{figure*}[t]

\begin{subfigure}{\textwidth}
\centering
  \includegraphics[clip,width=0.9\textwidth]{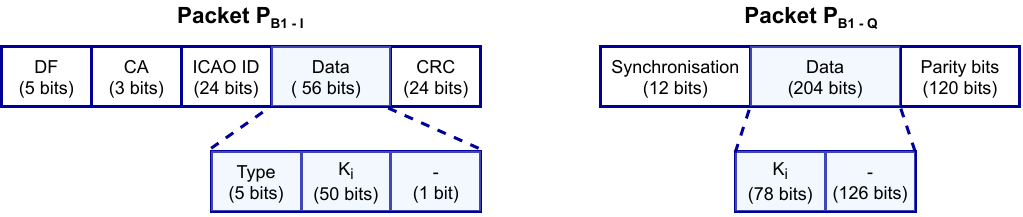}
  \caption{Structure of Type B1 packets, 112 bits for $P_{B1-I}$ and 336 bits $P_{B2-Q}$}\label{fig:PB1}
\end{subfigure}
\begin{subfigure}{\textwidth}
\centering
  \includegraphics[clip,width=0.9\textwidth]{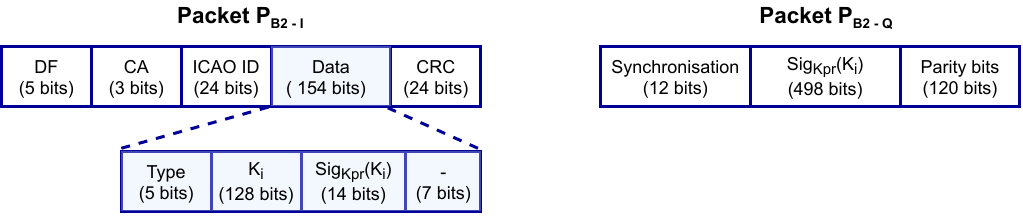}
  \caption{Structure of Type B2 packets, 210 bits for $P_{B2-I}$ and 630 bits $P_{B2-Q}$}\label{fig:PB2}
\end{subfigure}
\caption{Structure of Type B1 and B2 packets, conveying only the authentication key or the authentication key and its signature, respectively.}\label{fig:PB}
\end{figure*}

For Type B1 packets, the logical information $P_{B1}$ is split between packets $P_{B1-I}$ and $P_{B1-Q}$ that will be transmitted through the in-phase and quadrature components of the RF signal.  The in-phase packet $P_{B1-I}$ contains the 50 leftmost bits of the $K_I$ and the quadrature packet $P_{B1-Q}$ contains the remaining 78 bits, as indicated in Figure~\ref{fig:PB1}. 

For Type B2 packets, the information is similarly split into packets $P_{B2-I}$ and $P_{B2-Q}$.  The in-phase component $P_{B2-I}$ contains the entire interval key $K_i$ and the leftmost 14 bits of the signature, while the quadrature component $P_{B2-Q}$ contains the remaining 498 bits of the 512-bit signature.  

The signal components $S_{B1-I}$, $S_{B1_Q}$, $S_{B2-I}$ and $S_{B2-Q}$ are then generated similarly as for Type A packets (Equations~\ref{eq:PPMmod}, \ref{eq:DPSKmod} and \ref{eq:IQmod}).

\subsubsection{Sending the certificate of the transmitting aircraft}
Alice will broadcast the certificate of aircraft every $T_C$ seconds.  The Type C packet $P_C$ contains the public key of the aircraft $K_{pub}$ and the signature of this key $\sig_{K_{prCA}}(K_{pub})$.  With a security strength of 128 bits, an ECDSA public key size of 256 bits is required, resulting in a signature size of 512 bits \cite{FIPS186-5}.  The first 181 bits of the public key $K_{pub}$ are encoded in the in-phase packet $P_{C-I}$ and the remaining 75 bits at the beginning of the quadrature packet $P_{C-Q}$.  The 512 bits of the signature $\sig$ are also encoded into $P_{C-Q}$.

After encoding $P_{C-I}$ and $P_{C-Q}$, the sender generates the signals $S_{ C-I}$, $S_{C-Q}$, and finally the signal $S_C$ which she broadcast.  The procedure for producing these signals is the same as for producing signals for Type A and B packets.

\begin{figure*}[t]
\centering
\includegraphics[width=0.9\textwidth]{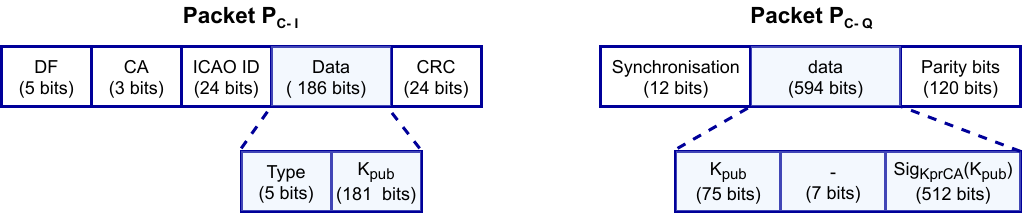}
\caption{Structure of Type C packets.  They contain the public key $K_{pub}$ of the aircraft and the signature of said key $\sig_{K_{prCA}}(K_{pub})$.  These packets are then used to generate the transmitted signal $S_C$.}\label{fig:PC}
\end{figure*}

\subsection{CABBA on the receiver side}\label{sec:CABBA-rcv}
\begin{figure}[h] 
\centering
\includegraphics[width=0.8\textwidth]{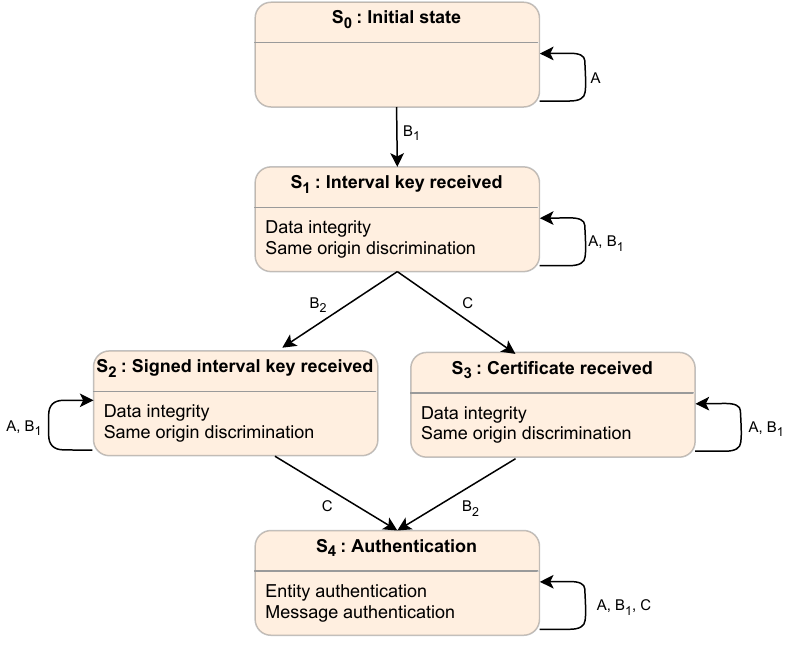}
\caption{State diagram illustrating the authentication process for ADS-B messages on the receiver side.}
\label{fig:Methodology}
\end{figure}

\subsubsection{Reception and demodulation of signals}
The process of receiving and demodulating messages by Bob (the receiver) is the same for all packet types.  The received signal $S$ is first demodulated with quadrature local oscillators to obtain the in-phase component $S_I$ and the quadrature component $S_Q$.  Bob then performs PPM demodulation onto $S_I$ and D8PSK demodulation onto $S_Q$, to generate logical packets $P_I$ and $P_Q$,
respectively.  Depending on the format of these packets, Bob determines the original packet type (A, B1, B2 or C) and processes them accordingly.  The processing of the CABBA messages contained within these logical packets is detailed below and depicted in the state diagram in Figure~\ref{fig:Methodology}.

\subsubsection{Processing the ADS-B message and its security data}
As described above, the ADS-B message information and its security data is contained in Type A packet $P_{A-I}$ and $P_{A-Q}$, respectively.  Bob thus extracts the message $m'$ from $P_{A-I}$ and the security data $\sigma$ (its sequence number $s$ and the MAC) from $P_{A-Q}$.  Lastly, he stores the triplet $(m',s,\mac)$ in a buffer until he can verify the integrity of the message.

\subsubsection{Verification of security properties}

CABBA is an asynchronous protocol and there is no guarantee that messages corresponding to a particular aircraft will be received in any particular order. The state diagram in Figure~\ref{fig:Methodology} describes the various states in which the receiver of CABBA could be depending on what security information, i.e.~what CABBA packet types, have been received so far.  Note that such a state diagram is used for all messages received with the same ICAO ID, i.e.~purportedly corresponding to the same aircraft.

% NAMES OF STATES
% S_0: Initial state
% S_1: Interval key received
% S_2: Signed interval key received
% S_3: Certificate received
% S_4: Authentication

The state machine is initialized at state $S_0$ when the first packet for a given ICAO ID is received.  If it is a Type A packet, it will be stored and the machine stays in the same state.  Reception of type B1 packet containing an interval key will generate a transition to State $S_1$.  Reception of a Type B2 packet, a \emph{signed} interval key,  will make the state machine transition to State $S_2$.  Finally, the (unlikely but possible) reception of a certificate in a Type C packet before a Type B1 or B2 packet will transition to State $S_3$. In all of these states ($S_1$, $S_2$ and $S_3$), subsequent reception of ADS-B messages in Type A packets and further interval keys in Type B packets causes no transitions.

At this point, the receiver is unable to perform either entity authentication or data origin authentication of any messages received because some security information is missing (has not been received), i.e.~either a validate certificate in the case of State $S_2$, a signed interval key in the case of State $S_3$, or both in State $S_1$.  Nonetheless, the receiver is able to perform data integrity verification of the ADS-B messages received in previous intervals.

% Data integrity verification
\paragraph{Data Integrity}
In order to validate the message integrity of a message $m'$ received during interval $i$, Bob must have already received Type B1 packets $P_{B1-I}$ and $P_{B1_Q}$ at the beginning of the next interval $i+1$.  From these packets, he will be able to reconstruct the interval key $K_i$ by concatenating the first 50 bits contained in $P_{B1-I}$ and the remaining 78 bits contained in $P_{B1-Q}$
(as shown in Figure~\ref{fig:PA}).  The next step is to calculate the authentication key $K_i'=F'(K_i)$.  Then, Bob calculates the ``correct'' HMAC of the received message $m'$ with this authentication key $K_i'$ as follows
$\hmac' = \hmac(m',K_i')$. Finally, Bob compares the $\lambda$ leftmost bits of $\hmac'$ with the received $\mac$.  If they coincide, Bob will accept the message $m'$, otherwise, he will ignore it.

Note that this verification only meets the goal of \emph{data integrity} of the message $m'$, i.e.~that the message has not been modified after its MAC was computed by its originator, whomever the originator might be (friend or foe, real aircraft or hacker).

%Same origin discrimination
\paragraph{Same-origin discrimination}
While in States $S_1$, $S_2$ and $S_3$ the receiver is unable to perform data
origin authentication, there is an important security property that can be asserted at
this point: \emph{same-origin discrimination}, i.e.~the ability to determine
which message was sent by which sender, without necessarily having authenticated
them.

To better understand this property, consider the following spoofing scenario.
The attacker is aware that Alice's aircraft with ICAO ID $X_A$ is currently
broadcasting ADS-B messages that Bob is receiving.  The intent of the spoofer is
to send counterfeit ADS-B messages bearing the same ICAO ID $X_A$.  Aware that the
aircraft ADS-B transmitter and the receiver have implemented CABBA, the spoofer
can generate his own interval key sequence and use it to authenticate his fake
messages.  More precisely, let $K^*=K^*_0,\ldots,K^*_N$ be the interval sequence
generated by the aircraft and let $K^\dag=K^\dag_0,\ldots,K^\dag_N$ be the
sequence generated by the spoofer.  The spoofer then generates his own messages,
including hashes computed with his own key sequence.

If both spoofer and Alice's aircraft are in Bob's reception range, Bob would
then receive these two message streams corresponding to the same ICAO ID $X_A$, potentially contradictory. Thus, while Bob may not be able to determine which messages came from the spoofer and which came from Alice's aircraft, he will still be able to correctly associate a new message $m'$ with either stream. He does this by identifying the TESLA keychain to which the interval key used to compute the MAC of $m'$ belongs, as described in Formula \ref{eq:sameOrigin}.

Consider two messages $m_1$ and $m_2$ received by Bob at intervals $i_1$ and $i_2$,
$i_2>i_1$, respectively, and whose integrity was verified by Bob in the
subsequent intervals $i_1+1$ and $i_2+1$ with interval keys $K_{i_1}$ and
$K_{i_2}$, respectively. Then if:
\begin{equation}\label{eq:sameOrigin}
K_{i_1}=F^{(i_2-i_1)}(K_{i_2})
\end{equation}
then Bob knows that $m_1$ and $m_2$ were sent by the same sender.
In other words, in the above scenario Bob will be able to detect
that there are two different senders $A^*$ and $A^\dag$ sending messages with
the same ICAO code, and further know which message corresponds to which sender.
He will not, however, know which one corresponds to the real aircraft $A$.

\paragraph{Authentication}
%Data origin authentication
When Bob has received all required security information, i.e.~a signed interval key and a certificate, he can then perform both identity authentication of the sender and message authentication (i.e.~data origin authentication) of previously received messages.  This will be possible when the state machine transitions to State $S_4$.\\

%Identity authentication
\textbf{Identity Authentication} Upon receiving Type C packet $P_{C-I}$ and $P_{C-Q}$, the public key $K_{\text{pub}}$ is extracted by concatenating the 181 bits in $P_{C-I}$ and the first 75 bits in $P_{C-Q}$.  The key signature $\sig_{K_{prCA}}(K_{pub})$ is
extracted from the remaining 512 bits of $P_{C-Q}$, as shown in Figure~\ref{fig:PC}.  Since Bob knows the public key $K_{pubCA}$ of the
Certificate Authority, he is able to verify the validity of the signature of the aircraft key the corresponding signature verification procedure \texttt{Verify}, that returns a boolean of \texttt{true} if the signature is valid.  In other
words, let $v_1$ be the result of this first signature verification:
\begin{equation} \label{eq:v1}
v_1= \texttt{Verify}(K_{pubCA},\sig_{K_{prCA}}(K_{pub}),K_{pub})
\end{equation}

%Interval key authentication
\textbf{Message authentication} Finally, having received a Type B2 packet $P_{B2-I}$ and $P_{B2-Q}$, Bob proceeds to extract the key $K_i$ from $P_{B2-I}$. To obtain the signature $\sig_{K_{pr}}(K_i)$, he concatenates the 14 bits from $P_{B2-I}$ with the 498 bits from $P_{B2-Q}$. Using this information, Bob verifies the authenticity of the interval key $K_i$. Let $v_2$ be the result of this verification procedure, as defined by the equation:
\begin{equation} \label{eq:v2}
v_2= \texttt{Verify}(K_{pub},\sig_{K_{pr}}(K_i),K_i)
\end{equation}
If $v_2$ evaluates \texttt{true}, it means that the key $K_i$ is authentic, i.e. that it has been generated by  Alice. Furthermore, this outcome also implies that all ADS-B messages for which the $\mac$ has been computed using $K_i$ can also be deemed authentic.

\section{Backward compatibility experiments} \label{section:experimentalP}

We conducted backward-compatibility tests to verify that the phase overlay capability, as implemented in CABBA, does not affect the ability of existing hardware to decode the original ADS-B messages that are transmitted in the in-phase component of the 1090ES carrier.

To do so, we built an SDR-based implementation of CABBA and tested its backward
compatibility with two distinct COTS ADS-B solutions:
\begin{enumerate}
\item The Appareo Stratus II ADS-B receiver, a non-certified portable device used in general aviation (GA) aircraft.  The Stratus II was connected via Wi-Fi to an IPad running the ForeFlight application displaying ADS-B traffic information.
\item The Collins TSS-4100, a certified avionics device integrating TCAS, transponder and ADS-B traffic surveillance capabilities, used in business jets and airline transport aircraft.  This equipment was connected to a Collins AFD-6520 Adaptative Flight Display to render the traffic information.
\end{enumerate}

The experimental setup we employed involved the following steps:
\begin{enumerate}
\item Generate ADS-B messages and corresponding CABBA packets using custom-made scripts.
\item Generate and transmit the corresponding RF signals using the HackRF One SDR.
\item Receive these RF signals with the corresponding COTS receiver.
\item Check that the transmitted ADS-B information is received and correctly interpreted.  We consider the test successful if the transmitted traffic information is displayed with the correct information (call sign, position, etc.).
\end{enumerate}

% Describe scripts
In our experimental setup, the ADS-B messages and the corresponding packets were generated using custom-made scripts.  These scripts are based on the \emph{ADSB\_Encoder.py} \cite{encoder.py} scripts.  This original script only generates ADS-B messages of the position report type, when given the ICAO, latitude, longitude, and altitude of an aircraft as inputs.  However, the logic of ADS-B receivers is such that in order for them to consider a given aircraft's traffic information, they must receive all required ADS-B message types, i.e.~identity, speed, status, and operating status, at the frequency prescribed by the protocol.  Thus, to conduct these tests, we built scripts that generate the remaining types of ADS-B messages\footnote{For this purpose the book \emph{The 1090 Megahertz Riddle} \cite{sun1090mhz} was an invaluable resource.}.

The scripts we constructed further added the functionality required to generate CABBA messages (keys and certificates) and the corresponding packets.  This includes among others functions to compute the MAC of ADS-B packets, to apply DPSK modulation to data to be transmitted in quadrature, to I-Q modulate the in-phase data in PPM with the quadrature data in DPSK.

For these backward-compatibility experiments, we only constructed and transmitted type A messages, which carry the ADS-B message and its security data.  We did not transmit the other types of CABBA messages (B1, B2 and C), as these would be ignored by legacy receivers since they do not contain ADS-B data.

\begin{figure}[ht] 
\centering
\includegraphics[width=0.98\textwidth]{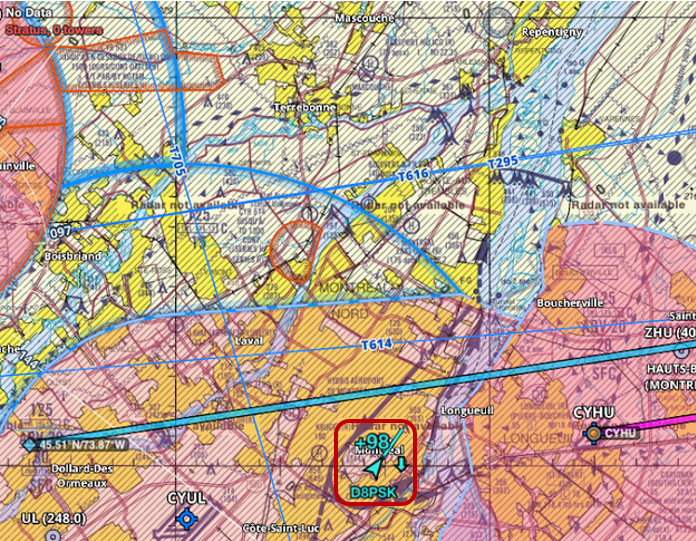}
\caption{Screen capture of the ForeFlight Maps display with traffic option activated, showing the correct information for the ``synthetic'' aircraft with call sign ``D8PSK'', obtained from an IPad connected to the Stratus II receiver.}
\label{fig:Stratus}
\end{figure}

\begin{figure}[ht] 
  \includegraphics[width=1\textwidth]{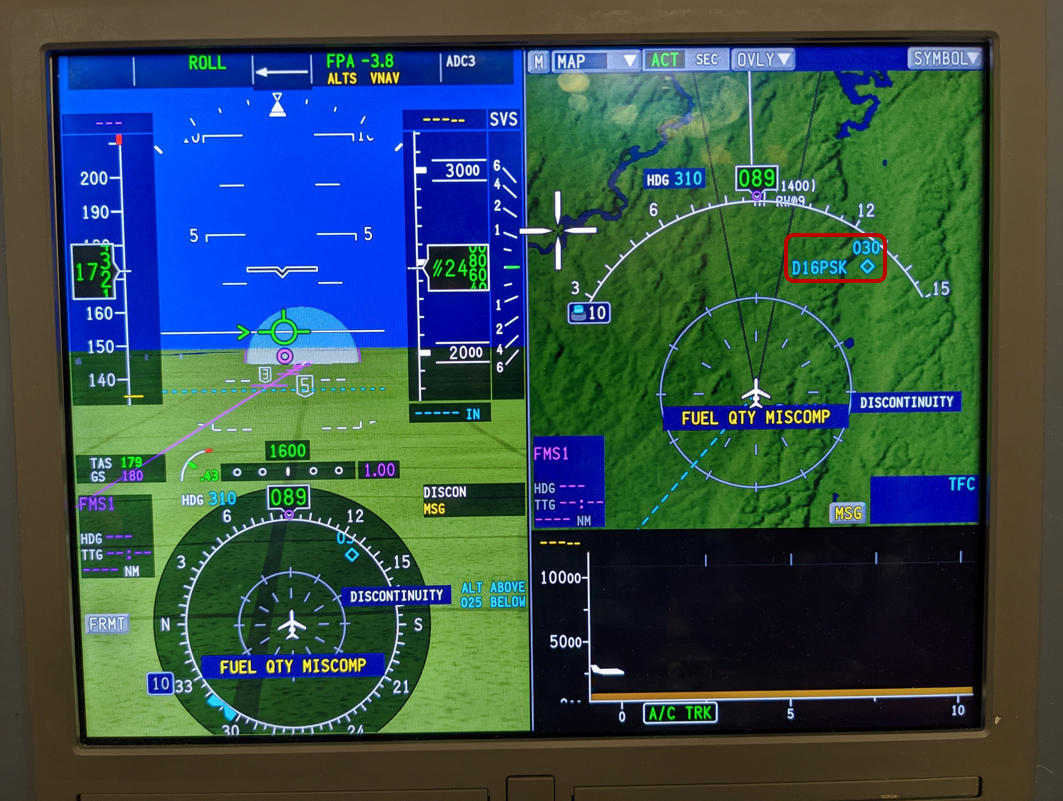}
\caption{Here is a picture of the display AFD-6520 Adaptive Flight Display showing the correct information for the ``synthetic'' aircraft with call sign ``D16PSK''.}
\label{fig:Collins}
\end{figure}

Our tests reveal that CABBA is backward-compatible with the two ADS-in receivers
under test.  Figure \ref{fig:Stratus} shows the display of the ForeFlight
application on the IPAD connected to the Stratus II receiver.  The information
displayed corresponds exactly with the information sent from the Hacker RF One
SDR.  The same is true for the information displayed on the AFD-6520 connected
to the TSS-4100 transponder, as shown in Figure~\ref{fig:Collins}.
The results suggest that using CABBA with legacy equipment will not compromise safety during the transitional period, where some aircraft would not yet have CABBA-capable ADS-B receivers. This finding supports the assumptions behind the MOPS used in CABBA and provides encouraging evidence for our implementation of it. However, further analysis and testing with a wider range of equipment in laboratory settings is needed, including packet reception analysis and, interoperability and stability tests. Once these tests are satisfactory, in-flight tests should follow, ideally in environments with high ADS-B channel usage and with sources of interference, such as multi-path transmissions due to terrain (mountains, water surface) or man-made obstacles (buildings, antennae, etc.).  While we do not think that the use of the MOPS would affect backward compatibility in such real-world conditions, we do believe that it is important to study how transmission and bit error rate for the quadrature signal would be affected by such sources of interference and in high-channel usage.

\section{Operational Viability of CABBA}
\label{section:experimentalP2}
While CABBA as proposed could provide a high level of security in terms of message authentication, there are some open questions regarding the viability of employing it in real-world situations due to operational and technological constraints.  

First, we must determine which modulation scheme is most appropriate, D8PSK or D16PSK.  Second, we must evaluate the bandwidth overhead of CABBA.  Even with the use of PSK modulation, CABBA still requires more transmission time than plain ADS-B.  It thus remains to be seen whether the resulting bandwidth overhead
challenges its use in the already congested 1~090~MHz frequency.  In this section, we describe our preliminary analysis of these questions using simulations to evaluate BER and real-world ADS-B data to conduct COR analysis.

\subsection{Comparative BER analysis of CABBA with D8PSK vs.\ D16PSK}
The aim of this BER analysis, is to determine which of the two phase overlay modulation schemes, D8PSK or D16PSK, provides the best balance between higher data throughput and acceptable signal quality, i.e.~ADS-B service quality.

We used Simulink \cite{matlab} to model the communication link of the CABBA protocol.  Then, we used the MATLAB program \texttt{bertool} to perform Monte-Carlo simulations to determine the BER across an Additive White Gaussian Noise (AWGN) channel.  

\begin{figure}[ht] 
\centering
\includegraphics[width=\textwidth]{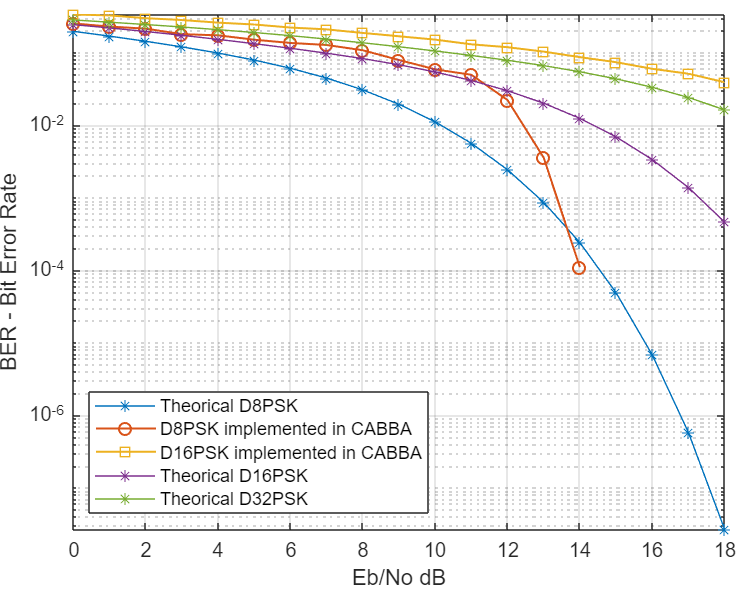}
\caption{BER analysis shows that CABBA meets the standard's requirements for
  normalized signal-to-noise values (Eb/No) greater than or equal to 15 dB when
  implemented with D8PSK.  Indeed, the standard requires a BER $< 10^{-6}$ and from
  Eb/No=15 the BER is equal to 0.}
\label{fig:BER}
\end{figure}

A lower BER indicates a better performance; for ADS-B, the standard establishes
a maximum BER of $10^-6$ \cite{RTCA260C2020}.  The BER curves of the two
implementations of CABBA that we wanted to compare, as well as the BER curves of
the D8PSK, D16PSK, and D32PSK modulations, are depicted in Figure~\ref{fig:BER}.
By observing these curves, we notice that:
\begin{enumerate}
   % \item As expected, CABBA performs better when implemented with D8PSK than when done with D16PSK.
    \item When implemented with D8PSK, CABBA fulfills the requirements of the
      standard for normalized signal-to-noise values (Eb/No) greater than or
      equal to 15 dB.  For these values, the BER is equal to zero, indicating
      that the transmission is error-free.
    \item When implemented using D16PSK, CABBA fails to meet the requirement of
      the standard.
\end{enumerate}

Based on these results, we find that the D8PSK technique is the best method for
implementing phase overlay functionality in avionics systems operating in the
1090ES band.  In the ADS-B context, the D16PSK technique has a significant
impact on data quality and reliability.  Given the high error rates provided by
D16PSK, the increase in throughput may not be worth it.

\subsection{Channel occupancy rate (COR) analysis}
We conducted a COR analysis to determine to what extent the transmission of
non-standard ADS-B information, which are essential for CABBA support, decreases
the available bandwidth.

The ITU report ITU-R SM.2256-1 \cite{ITU} provides a detailed discussion on
different approaches for measuring and evaluating spectrum occupancy, i.e.~a
methodology to conduct COR analyses.  We used it as a guide to conduct our
analysis.  Indeed, the activity factor ($\gamma$) reflects how active the
communication channel is. It is defined as follows \cite{sun2020analyzing}:
  \begin{equation} \label{eq:activityFactor}
    \gamma = \frac{\sum_{i=1}^{n} \Delta t_i}{\Delta t} 
  \end{equation}

where $\Delta t_i$ represents the channel occupation time for the $i$-th active
transmission and $\Delta t$ represents the total duration of the period being
considered.

We created a baseline of normal 1090ES channel occupancy levels using real ADS-B
data retrieved from the OpenSky Network database \cite{dataset1}. Since 2013, the OpenSky Network has been gathering continuous air traffic surveillance data as a non-profit community-based receiver network \cite{Schaefer14a}.  All unfiltered raw data is kept by OpenSky and made available to academic and institutional researchers.

To collect data for our research, we chose a receiver near Paris Orly airport
(IATA code \texttt{ORY}, ICAO code \texttt{LFPO}).  We chose this receiver because of the high
density of aircraft traffic that can come within its reception range, including:
\begin{enumerate}
\item Aircraft transiting through the Northern France airspace, i.e.~Paris Area
  Control Center (ACC), one of the busiest aerial corridors in the World.  The
  ADS-B station could receive signals from aircraft at cruise altitude
  (30-35,000 feet) up to 200 nautical miles (360 km).
\item Aircraft transiting through the Paris Terminal Maneuvering Area (TMA)
  that are landing or departing from Paris Charles de Gaulle, Orly or Le
  Bourget, some of the busiest airports in Europe.
  \item Aircraft on the ground at the Orly airport taxiing with transponders
    on.
\end{enumerate}
We obtained a data capture of all traffic for this station for a 24-hour period
on 3 August 2023.  Obviously, aircraft traffic varies during the day, and hence
so does 1090ES transmissions.  We sampled the traffic within each 1-hour period
and observed the transmission rate within 30 second-long periods within that
hour.  Taking six such samples for every hour, we observe quite a bit of
variation in the number of transmissions within each hour; the corresponding
confidence intervals are included in our results below.
\begin{figure}[h]
  \includegraphics[width=1\textwidth]{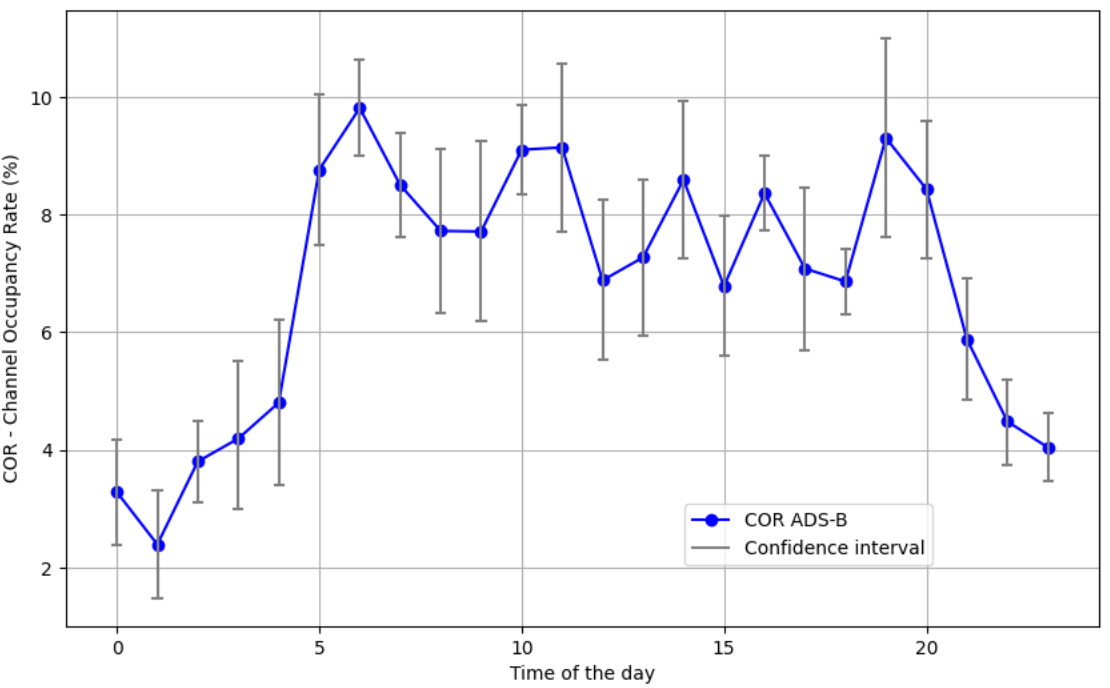}
  \caption{Mean COR for ADS-B transmissions with confidence intervals, computed
    with a sampling of six 30-second periods for every hour of the day, on 3
    August 2023.}
  \label{fig:COR1}
\end{figure}
\begin{table}
%\begin{tabular}{r|cccc}
  \centering
  \begin{tabular}{l||c|c|c|c}
           & Scenario 1 & Scenario 2 & Scenario 3 & Scenario 4 \\
  \hline
  $T_{B1}$ & 5 s    & 5 s     & 5 s   & 5 s \\
  $T_{B2}$ & 5 s    & 10 s   & 10 s   & 15 s \\
  $T_C$   & 5 s    & 15 s    & 20 s   & 30 s
  \end{tabular}
  \caption{Transmission period parameters for each of the four scenarios for
    which we computed the COR values.}
  \label{tab:CABBAparameters}
\end{table}
\begin{figure}[h]
  \includegraphics[width=1\textwidth]{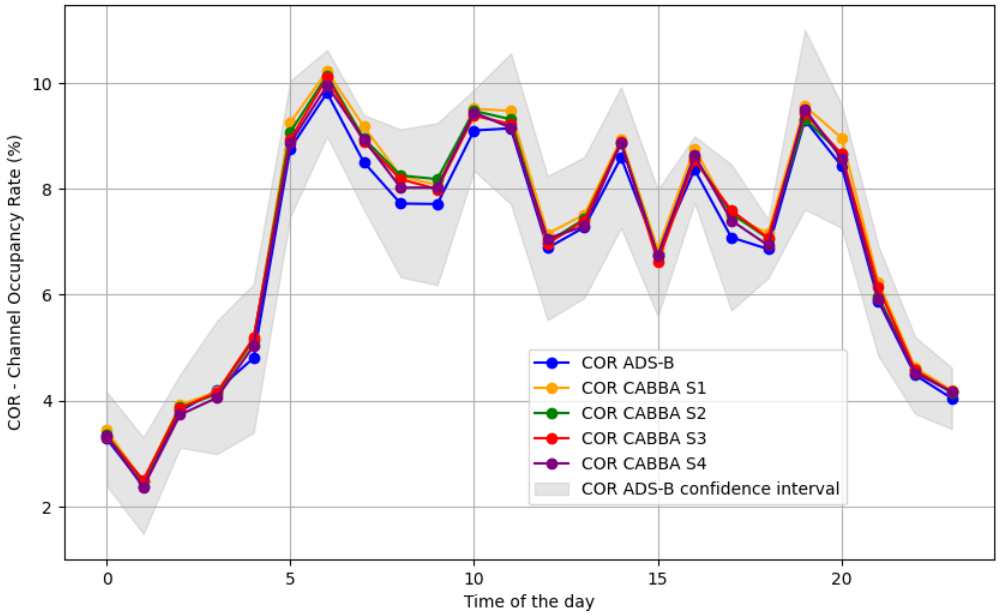}
  \caption{Estimated mean COR values per hour for every hour of 3 August 2023,
    for ADS-B and for CABBA in each of the three possible parameter settings
    described in Table~\ref{tab:CABBAparameters}.}
  \label{fig:COR2}
\end{figure}

\medskip
In CABBA, we transmit four different types of packets, i.e.~packet types A, B1,
B2 and C.  All packets of the same type have the same length and occupy the
channel for the same duration.  Let $\Delta t_A$, $\Delta t_{B1}$, $\Delta
t_{B2}$ and $\Delta t_C$ be the transmission times for each of these packet
types.  These values are proportional to the bit length of these packets
(Table~\ref{tab:CABBApacketTypes}) \emph{plus} the fixed 8-bit preamble.  Given
the 1090ES channel bit rate of 1 megabit/s, this results in 120 ms, 120 ms,
218~ms and 250 ms, respectively.

For each sampled time interval, let $n_A$, $n_{B1}$, $n_{B2}$ and $n_C$ be the
number of packets of each that would be transmitted with CABBA.  The resulting COR value is given by
\begin{equation}
  \gamma = \frac{n_A \, \Delta t_A + n_{B1}\, \Delta t_{B1} + n_{B2} \, \Delta
    t_{B2} + n_C \, \Delta t_C}{\Delta t}
  \label{eq:COR-CABBA}
\end{equation}

For our analysis, we conservatively consider that all aircraft in the dataset
are CABBA-capable and are sending all CABBA packet types as described in the
protocol description in Section~\ref{section:architecture}.  Of course, our
dataset only includes standard ADS-B packets and does not provide us with packet
counts for non-standard CABBA packet type, except for Type A packets for which
the count number $n_A$ is the same as the number of ADS-B packets.  For the
other packet types, we estimate the number of transmissions to be equal to
the number of different aircraft seen in the previous $T$ seconds \cite{Kistan2017}, where $T$ is
the transmission period of that type of packet.
For example, let us consider Type B1 packets, i.e.~packets containing unsigned
interval keys.  Each aircraft will send such a packet every $T_{B1}$ seconds,
i.e.~with a $T_{B1}$-second period.  At a given time $t$, let $x$ be the number
of different aircraft (with different ICAO ID) we have seen in our data set in
the previous $T_{B1}$ seconds.  During that time period, some aircraft will have
arrived in range and some others will have departed.  If we assume that within
the sampling interval (30 seconds) both the arrival and departure rate of
aircraft are relatively small and similar to each other, we then can safely
approximate the number of Type B1 packets that will be sent during that period
to be $x$, i.e.~$n_{B1}\approx x$.  The same can be said for the counts $n_{B2}$
of Type B2 and $n_C$ of Type C packets which we approximate to be the number of
different ICAO ID received in the previous $T_{B2}$ and $T_C$ seconds,
respectively.

With these approximations, we are then able to compute the COR value $\gamma$
from Equation~\ref{eq:COR-CABBA}.  We consider four different parameter settings,
described in Table~\ref{tab:CABBAparameters}.  In all scenarios, we keep the same
5-second Tesla interval duration set in SAT.  In the first scenario, Type C
packets are sent at every Tesla interval along with Type B2 packets; Type B1
packets are thus never sent.  This is the ``safest'' scenario in which
CABBA-compatible receivers must wait at most 5 seconds until being able to
authenticate messages, in the sense that the uncertainty period for CABBA
receivers where originators cannot be authenticated is minimized.  The fourth
scenario is the most bandwidth efficient, with Type B2 packets being sent every
other Tesla interval and Type C packet every six intervals.

% Results
%\subsection{Results and Interpretation}
The first finding of this study is that COR values for standard ADS-B vary
between 2.4\% and 9.8\%, for the quietest and busiest hours, 01h00 and 06h00
UTC, i.e.~03h00 and 08h00 Paris local time, respectively. 
Second, with respect to CABBA we observe that the maximum overhead corresponds,
obviously, to Scenario 1, with the lowest interarrival times for Type B1, B2 and
C packets.  In Scenario 1, the average overhead in terms of packets transmitted
is 3.76\% per 30-second period, with an observed maximum of 5.76\% during that
day.  In comparison, for the most bandwidth-efficient Scenario~4, the average
overhead is 1.56\% and maximised at 2.38\%.  With respect to channel occupancy
levels observed on that day, this results in a maximum COR increase of 0.43\% of
total bandwidth capacity for the most bandwidth-consuming Scenario 4.  These results are represented graphically in Figure~\ref{fig:COR2}.

In summary, whatever the parameter setting scenario we choose, the
impact of implementing CABBA in terms of bandwidth is negligible.  The
overhead in terms of packets sent is capped at less than 6\%.  This is
in sharp contrast with the original SAT proposal, which has an
estimated overhead of 35\%.

At the observed channel occupation levels, i.e.~COR values between 2\%
and 10\%, the COR increase for CABBA even its most bandwidth-consuming
setting is less than 1\%.  Even in more congested airspaces with
hypothetical COR values of up to 40\%, implementing CABBA would
increase COR to less than 43\%, a very acceptable compromise.

\subsection{Safety Impact of Unauthenticated Messages}
The key idea in TESLA that enables authenticated broadcast is the \emph{delayed} key disclosure.  However, this feature introduces a safety trade-off: messages cannot be immediately authenticated, resulting in an \emph{uncertainty delay} $\Delta_u$ between message arrival and authentication.

As discussed in Section~\ref{sec:CABBA-rcv}, to authenticate a message the following information must be in possession of the receiver: 1) the interval key for that message, 2) a signed interval key from this or a previous interval, and 3) the sender's certificate.  This information may have already been received in Type B1, B2 or C packets.   While the delay of arrival of such information is bounded by their respective interarrival periods, their order of arrival at the receiver is non-deterministic and thus the actual value of $\Delta_u$  is also non-deterministic.

Further analysis of what these uncertainty delays are and what is their impact on aviation safety depends on in what application and context the ADS-B information is being used.  We analyze various scenarios in TCAS and ATC applications below.

\subsubsection{Impact of packet loss on uncertainty delay}
In addition, we have to consider packet loss from noise, interference,
or collisions due to congestion, which could introduce a further delay
due to having to wait an additional 1 or~2 subsequent periods.  The
following equation describes the expected value of $\Delta_u$ as a function of the probability of packet loss $p$ and the interarrival period $T$ of the required information, where the first term corresponds to the packet being successfully received in the first interval, the second term in the second interval, etc.%

\begin{eqnarray}\label{delta-u}
  \overline{\Delta_u} & = & (1 - p - p^2) \frac{T}{2} + p  \left( T + \frac{T}{2} \right) +
  p^2 \left(2T + \frac{T}{2} \right) \nonumber \\
  & = & \frac{T}{2} (1+2p+4p^2)
\end{eqnarray}

Previous works \cite{Schaefer14a,sciancalepore2019reliability} have explored the packet loss ratio for ADS-B. They apply a conservative definition, where packet loss occurs if \emph{any} receiver within the aircraft's range is missing the corresponding reception record.  In the rest of this section, we approximate the probability of packet loss at a particular receptor, by interpolating from the empirical cumulative distribution function between packet loss and distance in \cite{sciancalepore2019reliability}.

\subsubsection{Uncertainty Delays in TCAS}
TCAS technology is essential in two environments: 1) non-radar environments, like oceanic regions, where ATC may be unable to provide separation services, and 2) dense traffic areas, such as busy airport terminals, where separation conflicts are likely to occur.  The TCAS standards and supporting equipment ensure pilots have sufficient situational awareness of nearby aircraft within predefined \emph{protection volumes}.  These volumes are defined to address potential separation conflicts within specific \emph{time} periods: 20--48 seconds for traffic advisories (TA) and 15--35 seconds for resolution advisories (RA) \cite{TCAS2}. These intervals provide pilots with adequate time to enhance their situational awareness of incoming traffic (TA) and to execute evasive maneuvers (RA). It is thus very important that uncertainty delays do not significantly reduce the reaction time for pilots.

For aircraft in flight, the line-of-sight range corresponds to the distance to the horizon.  For typical altitudes that are small in comparison with the radius of the Earth, this can be approximated as follows:
  \begin{equation}
    \text{LOS range (in NM)}=1.06*\sqrt{\text{altitude (in feet)}}
  \end{equation}
In remote and oceanic areas, TCAS establishes a maximum lateral closure rate of 1~200 knots (1~200 nautical miles (NM) per hour = 2~222 km/h) \cite{TCAS2}, resulting in protected volumes with radii of 16 NM for TA and 11,6 NM for RA (distances corresponding to this closure velocity and the specified time periods for TA and RA).  At a cruise altitude of 35~000 feet, the line-of-sight (LOS) range between aircraft is approximately 396,6 NM, providing 19,8 minutes of transmission time before the aircraft enters the protected volume.  In terminal areas, arriving aircraft typically operate between 10~000 and 3~000 feet, resulting in a worst-case minimum LOS range of 116,1 NM (both at 3~000 feet).  With a speed limit of 250 knots and a maximum closure rate of 500 knots, these aircraft will enter LOS range at least 13,9 minutes before entering each other's protected volumes. 
In both cases, this is significantly larger than the proposed $T_{B2}$ and $T_C$ periods, which guarantees reception of the required Type B1 and C packets, with high probability, even accounting for the packet loss rates at those distances.  Thus, the only significant delay to be considered is the one due to the delayed transmission of the interval keys ($T_{B1}$).

At distances commensurate with the radii of protected volumes, the probability of packet loss for Type B1 packets is modest and the resulting expected value of $\Delta_u$ is less than 3,0 s, for both TA and RA.  In other words, the use of CABBA reduces reaction time by at most 3 s, which we deem acceptable compared with the above-mentioned overall reaction times for TA and RA protected volumes.

\begin{table}[ht]
    \hskip-1.7cm
    \begin{tabular}{|c||c|c|c|c||c|c|}
      \hline
      \textbf{TCAS} & \textbf{radius} & $p$ (\%) & $\Delta_u(T_{B1})$ (s) & \textbf{time} (s)& \textbf{LOS} (min) & $\Delta_u(T_C)$ (s)\\
      \hline
        TA & 6,6--16 NM  & 8,9 & 3,0 & 20--48   & 13,9--19,8 & 18,1 \\
           & (12,2--29,6 Km) & & & & & \\
        \hline
        RA & 5--11,6 NM  & 6,4 & 2,9& 15--35 & 13,9--19,8 & 17,2 \\
           & (9,3--21,5 Km) & & & & & \\
        \hline
    \end{tabular}
    \caption{Packet loss probability and uncertainty times for TCAS.  We give the \emph{radius} for the corresponding protected volume, the derived probability $p$ of packet loss at that distance (as per \cite{sciancalepore2019reliability}), the resulting expected authentication delay due to interval key transmission $\Delta_u(T_{B1})$, and compare it with the total reaction time as per the TCAS standard \cite{TCAS2}.  In addition, we compare the time in the LOS range with the expected uncertainty delay for transmission of Type C packets $\Delta_u(T_c)$.  Here, we use the parameters for the ``worst-case'' Scenario 4 of Table~\ref{tab:CABBAparameters}, i.e.~$T_{B1}=5$ s and $T_{B2}=T_C=30$ s.}
    \label{tab:packetloss}
\end{table}

\subsubsection{Uncertainty Delays in ATC}
In ATC, air traffic controllers utilize interconnected air traffic management (ATM) systems that can communicate with airborne aircraft via digital channels (e.g.~ACARS, CPDLC). It is reasonable to expect that these systems will have access to a PKI with aircraft certificates or receive them along with flight plan information.  The same can be said of signed keys for the initial interval of a flight, that could be transmitted by the aircraft at the gate.  In this case, the primary factor affecting uncertainty delays would be the interarrival time of interval keys, i.e.~the parameter $T_{B1}$.

If the certificate or signed key is unavailable, then the ATM system may have to rely on their transmission by CABBA through Type B1 and C2 packets, with the corresponding authentication uncertainty delay.  The impact of this uncertainty will depend on the type and size of airspace being controlled.
  \begin{description}
  \item[Airport control zones (aka ``Tower'').] A sector centered on an airport, with a typical radius of 5 NM and a maximum altitude of 3~000 feet.
  \item[Terminal areas.] A inverted-cone shaped sector above the control zone, with up to 40 NM in radius and up to 12~500 feet.
    \item[Area Control Center.] A large sector, covered by surveillance from a single surveillance installation (radar or ADS-B receiver), typically with a 100--150 NM radius, to ensure that all aircraft in the area are well within minimum LOS range (approx.~140 NM for typical cruise altitudes18--60~000 feet in class A controlled airspace).  
  \end{description}

For tower and terminal areas, the LOS range is 58 and 118,5 NM, respectively.  With a maximum speed of 250 knots, this results in minimum time in LOS for approaching aircraft of 13,9 and 28,4 min, respectively.  For ACC, the LOS range is 140 NM for aircraft in Class A airspace, which at an unrestricted max cruise speed of 450 knots, results in a minimum time in LOS range of 18,6 min. As was the case with TCAS, in all of these cases, these LOS range transition times are widely sufficient for the ADS-B receiver to receive Type B2 and C packets with high probability, even considering the packet loss probabilities for those distances, as shown in Table~\ref{tab:ATC}.

As for the uncertainty delay created by the reception of the interval keys, in the case of tower and terminal areas, these are 3,0 and 4,1 s respectively, which compares very favorably with the refresh rate of typical Secondary Surveillance Radar (SSR) installations used for ATC (6 antenna rotations per minute, resulting a 10 s refresh period).  The worst case here is that of the ACC, where the delay can be up to 14 sec, which remains comparable to the delay found in traditional radar-based infrastructures.

\begin{table}[ht]
    \hskip-1.7cm
    \begin{tabular}{|c||c|c|c|c||c|c|}%{|l|l|c|c|c|}
      \hline
      \textbf{ATC} & \textbf{radius} & $p$ (\%) & $\Delta_u(T_{B1})$ (s) & \textbf{update} (s) & \textbf{LOS} (min) &$\Delta_u(T_C)$ (s)\\
      \hline
        Tower & 5 NM     & 2,8 & 3,0 & 10 & 13,9 & 16,3  \\
              & (9,3 Km) &     &     &    &      &       \\
        \hline
        Terminal & 5--40 NM       & 22,2 & 4,1 & 10 & 28,4 & 25,0 \\
                 & (9,3--74,1 Km) &      &     &    &      & \\
        \hline
        ACC & 100--150 NM       & 83,3 & 14,0 & 10 & 18,6 & 82,1 \\
            & (185,2--277,8 Km) &      &      &    &      & \\
        \hline
    \end{tabular}
    \caption{Packet loss probability and uncertainty times for ATC.
      We give the \emph{radius} for the corresponding control area,
      the derived probability $p$ of packet loss, the expected
      authentication delay for interval key reception
      $\Delta_u(T_{B1})$, and compare it with the typical radar update
      rate.  We also compare the time in the LOS range with the
      expected uncertainty delay for Type C packet reception
      $\Delta_u(T_c)$.  We use the same parameters as before,
      i.e.~$T_{B1}=5$ s and $T_{B2}=T_C=30$ s.}
    \label{tab:ATC}
\end{table}

\section{Conclusion} \label{section:conclusion}

In this paper, we have explored the Compatible Authenticated Bandwidth-efficient Broadcast protocol for ADS-B (CABBA), a proposal designed to secure the Automatic Dependent Surveillance - Broadcast (ADS-B) protocol used in aviation. CABBA integrates the TESLA authentication protocol into the application layer of ADS-B. It also incorporates the phase overlay modulation techniques outlined in the Minimum Operational Performance Standard (MOPS) \cite{RTCA260C2020} into the physical layer of ADS-B. With these enhancements, CABBA strengthens ADS-B security and ensures the safety of the protocol by complying with the rigorous operational standards set by MOPS.

From an \emph{operational and technical} point of view, CABBA  shows promise. On one hand, preliminary tests indicate that the use of phase-overlay modulation techniques (D8PSK) proposed in the MOPS does not affect the capacity of legacy receivers to correctly interpret ADS-B messages.  This would enable CABBA-compliant ADS-B hardware to co-exist with legacy ADS-B equipment without compromising safety. On the other hand, simulations using real ADS-B traffic data from high-traffic environments suggest a tolerable channel occupancy rate overhead when deploying CABBA. The results indicate that the bandwidth overhead when using CABBA is very reasonable and should not impede its deployment, even in congested airspace. Moreover, given the hardware and software architecture of most modern avionics systems, transforming legacy ADS-B equipment to support CABBA could probably be done with a firmware and software upgrade (e.g.~in avionics using FPGA for signal
processing).  In such situations, the cost of upgrades and time to availability
and certification would be less than a full avionics replacement.

%PKI
From an \emph{organizational} point of view, however, a robust international public-key infrastructure (PKI) must be established and operated. While there are ICAO ID databases in operations, they do not currently support certificate-based PKI sharing of aircraft public keys. Implementing such a PKI would require global agreement on trustworthy organizations to manage, share, and store aircraft public keys and their certificates. Although this poses a significant challenge, we believe it is doable in the relatively short term. A similar infrastructure exists for the sharing and storage of public keys for electronic Machine Readable Travel Documents (eMRTDs), including biometric passports \cite{ICAOPKD}. It is currently supported by the ICAO Public Key Directory (PKD) with 90 participating countries \cite{ICAOPKDparticipants}. We believe this framework could be expanded to include aircraft certificates for authenticating CABBA messages.

In conclusion, for all of these reasons supported by our experimental work and
analysis, we believe that CABBA offers the best choice for a quicker deployment
of a secure ADS-B solution that meets operational and technological requirements, while simultaneously achieving security and aviation safety objectives.

\section*{Acknowledgment}
The authors would like to thank the partners of the CyberSA project: Queen's University, Bombardier, Collins Aerospace, Rhea Group, Carillon Information Security, the International Air Transport Association (IATA), the Natural Sciences and Engineering Research Council of Canada (NSERC) and the Consortium for Research and Innovation in Aerospace in Québec (CRIAQ). Their support has been pivotal in bringing this research to fruition. We sincerely appreciate their collaborative efforts to advance research in the field of avionics systems cybersecurity.
\section*{Declaration of generative AI and AI-assisted technologies in the writing process}
During the preparation of this work the author(s) used DeepL, Quillbot, ChatGPT, Grammarly and Turnitin in order to : 1) translate from French and Spanish to English (DeepL), 2) rephrase some sentences from previous work to write the SOTA (Quillbot and ChatGPT), 3) enhance the quality and clarity of specific sentences (ChatGPT), 4) check grammar and spelling mistakes (Grammarly), check for plagiarism (Turnitin). After using this tool/service, the author(s) reviewed and edited the content as needed and take(s) full responsibility for the content of the publication.

\bibliographystyle{IEEEtran}
\bibliography{mybibfile.bib}

% Generated by IEEEtran.bst, version: 1.14 (2015/08/26)
\begin{thebibliography}{10}
\providecommand{\url}[1]{#1}
\csname url@samestyle\endcsname
\providecommand{\newblock}{\relax}
\providecommand{\bibinfo}[2]{#2}
\providecommand{\BIBentrySTDinterwordspacing}{\spaceskip=0pt\relax}
\providecommand{\BIBentryALTinterwordstretchfactor}{4}
\providecommand{\BIBentryALTinterwordspacing}{\spaceskip=\fontdimen2\font plus
\BIBentryALTinterwordstretchfactor\fontdimen3\font minus \fontdimen4\font\relax}
\providecommand{\BIBforeignlanguage}[2]{{%
\expandafter\ifx\csname l@#1\endcsname\relax
\typeout{** WARNING: IEEEtran.bst: No hyphenation pattern has been}%
\typeout{** loaded for the language `#1'. Using the pattern for}%
\typeout{** the default language instead.}%
\else
\language=\csname l@#1\endcsname
\fi
#2}}
\providecommand{\BIBdecl}{\relax}
\BIBdecl

\bibitem{yang2022aircraft}
X.~Yang, J.~Sun, and R.~T. Rajan, ``Aircraft trajectory prediction using ads-b data,'' in \emph{Pre-Proceedings of the 2022 Symposium on Information Theory and Signal Processing in the Benelux}, 2022, p. 113.

\bibitem{adsbRTCA2012}
RTCA, ``{M}inimum {O}perational {P}erformance {A}utomatic {D}ependent {S}urveillance-{B}roadcast {(ADS-B)} and {T}raffic {I}nformation {S}ervices-{B}roadcast {(TIS-B)},'' Radio {T}echnical {C}ommission for {A}eronautics, Washington, DC, Technical report {DO-260B}, 2011.

\bibitem{RTCA260C2020}
------, ``{M}inimum {O}perational {P}erformance {S}tandards for 1090 {MHz} {E}xtended {S}quitter {A}utomatic {D}ependent {S}urveillance – {B}roadcast {(ADS-B)} and {T}raffic {I}nformation {S}ervices – {B}roadcast {(TIS-B)},'' Radio Technical Commission for Aeronautics, Washington, DC, Technical report DO-260C, 2020.

\bibitem{4444PANS}
ICAO, ``Doc 4444, {P}rocedures for {A}ir {N}avigation {S}ervices — {A}ir {T}raffic {M}anagement,'' International {C}ivil {A}viation {O}rganization, Montréal, QC, Technical report Doc 4444 PANS-ATM, 2016.

\bibitem{CNSDef}
S.~Thompson, D.~Spencer, and J.~Andrews, ``An {A}ssessment of the {C}ommunications, {N}avigation, {S}urveillance {(CNS)} {C}apabilities {N}eeded to {S}upport the {F}uture {A}ir {T}raffic {M}anagement {S}ystem,'' {M}assachusetts {I}nstitute of {T}echnology, {L}incoln {L}aboratory, Cambridge, MA, Technical report, 2001.

\bibitem{ED129}
EUROCAE, ``{T}echnical {S}pecification for the {ADS-B} {G}round {S}tation,'' European {O}rganisation for {C}ivil {A}viation {E}quipment, Paris, FR, Technical report ED-129, 2010.

\bibitem{ED129A}
------, ``{T}echnical {S}pecification for a 1090 {MHZ} {E}tended {S}quitter {ADS-B} ground station,'' European {O}rganisation for {C}ivil {A}viation {E}quipment, Paris, FR, Technical report ED-129, 2015.

\bibitem{NextGen}
\BIBentryALTinterwordspacing
``Next {G}eneration {A}ir {T}ransportation {S}ystem ({NextGen}),'' {Federal Aviation Administration (FAA)}, Jul. 2022. [Online]. Available: \url{https://www.faa.gov/nextgen}
\BIBentrySTDinterwordspacing

\bibitem{datalink4ADS-B}
\BIBentryALTinterwordspacing
{ICAO}, ``Surveillance of remotely piloted aircraft systems ({RPAS}) and cybersecurity,'' Technical Commission {R}ussian {F}ederation, Tech. Rep. A39-WP/296, 2016. [Online]. Available: \url{https://www.icao.int/Meetings/a39/Documents/WP/wp_296_en.pdf}
\BIBentrySTDinterwordspacing

\bibitem{costin2012ghost}
A.~Costin and A.~Francillon, ``Ghost in the air (traffic): On insecurity of ads-b protocol and practical attacks on ads-b devices,'' \emph{black hat USA}, vol.~1, pp. 1--12, 2012.

\bibitem{manesh2017analysis}
M.~R. Manesh and N.~Kaabouch, ``Analysis of vulnerabilities, attacks, countermeasures and overall risk of the automatic dependent surveillance-broadcast {(ADS-B)} system,'' \emph{International Journal of Critical Infrastructure Protection}, vol.~19, pp. 16--31, 2017.

\bibitem{strohmeier2014realities}
M.~Strohmeier, M.~Sch{\"a}fer, V.~Lenders, and I.~Martinovic, ``Realities and challenges of nextgen air traffic management: the case of ads-b,'' \emph{IEEE Communications Magazine}, vol.~52, no.~5, pp. 111--118, 2014.

\bibitem{ryon2018safety}
L.~Ryon and G.~Rice, ``A safety-focused security risk assessment of commercial aircraft avionics,'' in \emph{2018 IEEE/AIAA 37th Digital Avionics Systems Conference (DASC)}.\hskip 1em plus 0.5em minus 0.4em\relax IEEE, 2018, pp. 1--8.

\bibitem{perrig2002tesla}
A.~Perrig, R.~Canetti, J.~D. Tygar, and D.~Song, ``The {TESLA} broadcast authentication protocol,'' \emph{Rsa Cryptobytes}, vol.~5, no.~2, pp. 2--13, 2002.

\bibitem{perrig2003tesla}
A.~Perrig and J.~Tygar, ``Tesla broadcast authentication,'' in \emph{Secure Broadcast Communication}.\hskip 1em plus 0.5em minus 0.4em\relax Springer, 2003, pp. 29--53.

\bibitem{finke2013ads}
C.~Finke, J.~Butts, and R.~Mills, ``{ADS-B} encryption: confidentiality in the friendly skies,'' in \emph{Proceedings of the Eighth Annual Cyber Security and Information Intelligence Research Workshop}.\hskip 1em plus 0.5em minus 0.4em\relax ACM, 2013, pp. 1--4.

\bibitem{finke2013enhancing}
C.~Finke, J.~Butts, R.~Mills, and M.~Grimaila, ``Enhancing the security of aircraft surveillance in the next generation air traffic control system,'' \emph{International Journal of Critical Infrastructure Protection}, vol.~6, no.~1, pp. 3--11, 2013.

\bibitem{huang2014enabling}
R.~S. Huang, H.~M. Yang, and H.~G. Wu, ``Enabling confidentiality for {ADS-B} broadcast messages based on format-preserving encryption,'' in \emph{Applied Mechanics and Materials}, vol. 543.\hskip 1em plus 0.5em minus 0.4em\relax Trans Tech Publ, 2014, pp. 2032--2035.

\bibitem{abgeyibor2014evaluation}
R.~Agbeyibor, J.~Butts, M.~Grimaila, and R.~Mills, ``Evaluation of format-preserving encryption algorithms for critical infrastructure protection,'' in \emph{International Conference on Critical Infrastructure Protection}.\hskip 1em plus 0.5em minus 0.4em\relax Springer, 2014, pp. 245--261.

\bibitem{Habibi2023}
\BIBentryALTinterwordspacing
J.~Habibi~Markani, A.~Amrhar, J.-M. Gagné, and R.~J. Landry, ``Security establishment in ads-b by format-preserving encryption and blockchain schemes,'' \emph{Applied Sciences}, vol.~13, no.~5, 2023. [Online]. Available: \url{https://www.mdpi.com/2076-3417/13/5/3105}
\BIBentrySTDinterwordspacing

\bibitem{samuelson2006enhanced}
K.~Samuelson, E.~Valovage, and D.~Hall, ``Enhanced {ADS-B} research,'' in \emph{2006 IEEE Aerospace Conference}.\hskip 1em plus 0.5em minus 0.4em\relax IEEE, 2006, pp. 7--pp.

\bibitem{kacem2015integrity}
T.~Kacem, D.~Wijesekera, and P.~Costa, ``Integrity and authenticity of {ADS-B} broadcasts,'' in \emph{2015 IEEE Aerospace Conference}.\hskip 1em plus 0.5em minus 0.4em\relax IEEE, 2015, pp. 1--8.

\bibitem{feng2010data}
Z.~Feng, W.~Pan, and Y.~Wang, ``A data authentication solution of ads-b system based on x. 509 certificate,'' in \emph{27th International Congress of the Aeronautical Sciences, ICAS}, 2010, pp. 1--6.

\bibitem{buchholz2013dpp}
A.~K. Buchholz, ``Dpp: Dual path pki for secure aircraft data communication,'' Ph.D. dissertation, Virginia Polytechnic Institute and State University, 2013.

\bibitem{baek2013authentication}
J.~Baek, Y.-J. Byon, E.~Hableel, and M.~Al-Qutayri, ``An authentication framework for automatic dependent surveillance-broadcast based on online/offline identity-based signature,'' in \emph{2013 Eighth International Conference on P2P, Parallel, Grid, Cloud and Internet Computing}.\hskip 1em plus 0.5em minus 0.4em\relax IEEE, 2013, pp. 358--363.

\bibitem{yang2013efficient}
H.~Yang, H.~Kim, H.~Li, E.~Yoon, X.~Wang, and X.~Ding, ``An efficient broadcast authentication scheme with batch verification for ads-b messages,'' \emph{KSII Transactions on Internet and Information Systems (TIIS)}, vol.~7, no.~10, pp. 2544--2560, 2013.

\bibitem{yang2014ebaa}
H.~Yang, R.~Huang, X.~Wang, J.~Deng, and R.~Chen, ``Ebaa: An efficient broadcast authentication scheme for ads-b communication based on ibs-mr,'' \emph{Chinese Journal of Aeronautics}, vol.~27, no.~3, pp. 688--696, 2014.

\bibitem{yang2015new}
A.~Yang, X.~Tan, J.~Baek, and D.~S. Wong, ``A new ads-b authentication framework based on efficient hierarchical identity-based signature with batch verification,'' \emph{IEEE Transactions on Services Computing}, vol.~10, no.~2, pp. 165--175, 2015.

\bibitem{he2016efficient}
D.~He, N.~Kumar, K.-K.~R. Choo, and W.~Wu, ``Efficient hierarchical identity-based signature with batch verification for automatic dependent surveillance-broadcast system,'' \emph{IEEE Transactions on Information Forensics and Security}, vol.~12, no.~2, pp. 454--464, 2016.

\bibitem{thumbur2019efficient}
G.~Thumbur, N.~Gayathri, P.~V. Reddy, M.~Z.~U. Rahman \emph{et~al.}, ``Efficient pairing-free identity-based ads-b authentication scheme with batch verification,'' \emph{IEEE Transactions on Aerospace and Electronic Systems}, vol.~55, no.~5, pp. 2473--2486, 2019.

\bibitem{braeken2019holistic}
A.~Braeken, ``Holistic air protection scheme of {ADS-B} communication,'' \emph{IEEE Access}, vol.~7, pp. 65\,251--65\,262, 2019.

\bibitem{wu2019ads}
Z.~Wu, A.~Guo, M.~Yue, and L.~Liu, ``An {ADS-B} message authentication method based on certificateless short signature,'' \emph{IEEE Transactions on Aerospace and Electronic Systems}, vol.~56, no.~3, pp. 1742--1753, 2019.

\bibitem{asari2021new}
A.~Asari, M.~R. Alagheband, M.~Bayat, and M.~R. Asaar, ``A new provable hierarchical anonymous certificateless authentication protocol with aggregate verification in ads-b systems,'' \emph{Computer Networks}, vol. 185, p. 107599, 2021.

\bibitem{subramani2021efficient}
J.~Subramani, A.~Maria, R.~B. Neelakandan, and A.~S. Rajasekaran, ``Efficient anonymous authentication scheme for automatic dependent surveillance-broadcast system with batch verification,'' \emph{IET Communications}, vol.~15, no.~9, pp. 1187--1197, 2021.

\bibitem{yang2017lhcsas}
H.~Yang, M.~Yao, Z.~Xu, and B.~Liu, ``Lhcsas: a lightweight and highly-compatible solution for {ADS-B} security,'' in \emph{GLOBECOM 2017-2017 IEEE Global Communications Conference}.\hskip 1em plus 0.5em minus 0.4em\relax IEEE, 2017, pp. 1--7.

\bibitem{berthier2017sat}
P.~Berthier, J.~M. Fernandez, and J.-M. Robert, ``{SAT}: {S}ecurity in the {A}ir using {T}esla,'' in \emph{2017 IEEE/AIAA 36th Digital Avionics Systems Conference (DASC)}.\hskip 1em plus 0.5em minus 0.4em\relax IEEE, 2017, pp. 1--10.

\bibitem{sciancalepore2018sos}
S.~Sciancalepore and R.~Di~Pietro, ``{SOS}-{S}ecuring {O}pen {S}kies,'' in \emph{International Conference on Security, Privacy and Anonymity in Computation, Communication and Storage}.\hskip 1em plus 0.5em minus 0.4em\relax Springer, 2018, pp. 15--32.

\bibitem{NISSP80038GRev1}
NIST, ``{R}ecommendation for {B}lock {C}ipher {M}odes of {O}peration:\emph{{M}ethods for {F}ormat-{P}reserving {E}ncryption},'' National {I}nstitute of {S}tandards and {T}echnology {(NIST)}, Gaithersburg, MD, Technical Report NIST Special Publication (SP) 800-38G Rev. 1, 2019.

\bibitem{Challal2004MulticastAuth}
Y.~Challal, H.~Bettahar, and A.~Bouabdallah, ``A taxonomy of multicast data origin authentication: {I}ssues and solutions,'' \emph{IEEE Communications Surveys \& Tutorials}, vol.~6, no.~3, pp. 34--57, 2004.

\bibitem{DigitalSignatureStandard}
\BIBentryALTinterwordspacing
{NIST}, ``{D}igital {S}ignature {S}tandard {(DSS)},'' {U.S.} {D}epartment of {C}ommerce, Gaithersburg, MD, Standard FIPS PUB 186-4, 2013. [Online]. Available: \url{https://nvlpubs.nist.gov/nistpubs/FIPS/NIST.FIPS.186-4.pdf}
\BIBentrySTDinterwordspacing

\bibitem{NISSP800175BRev1}
NIST, ``{G}uideline for {U}sing {C}ryptographic {S}tandards in the {F}ederal {G}overnment: {C}ryptographic {M}echanisms,'' National {I}nstitute of {S}tandards and {T}echnology {(NIST)}, Gaithersburg, MD, Technical Report NIST Special Publication (SP) 800-175B Rev. 1, 2020.

\bibitem{FIPS186-5}
\BIBentryALTinterwordspacing
{NIST}, ``{D}igital {S}ignature {S}tandard {(DSS)},'' {U.S.} {D}epartment of {C}ommerce, Gaithersburg, MD, Standard FIPS PUB 186-5, 2023. [Online]. Available: \url{https://nvlpubs.nist.gov/nistpubs/FIPS/NIST.FIPS.186-5.pdf}
\BIBentrySTDinterwordspacing

\bibitem{EncyclopediaCryptoandSec2011}
\BIBentryALTinterwordspacing
H.~C. Van~Tilborg and S.~Jajodia, Eds., \emph{Encyclopedia of {C}ryptography and {S}ecurity}, 2nd~ed.\hskip 1em plus 0.5em minus 0.4em\relax {N}ew {Y}ork, {NY}: Springer {N}ew {Y}ork, 2011. [Online]. Available: \url{https://doi.org/10.1007/978-1-4419-5906-5}
\BIBentrySTDinterwordspacing

\bibitem{NISTSP80057}
\BIBentryALTinterwordspacing
{NIST}, ``{R}ecommendation for {K}ey {M}anagement: \emph{Part 2 – {B}est {P}ractices for {K}ey {M}anagement {O}rganizations},'' {U.S.} {D}epartment of {C}ommerce, Gaithersburg, MD, Standard NIST SP 800-57 PT . 2 R EV . 1, 2019. [Online]. Available: \url{https://nvlpubs.nist.gov/nistpubs/SpecialPublications/NIST.SP.800-57pt2r1.pdf}
\BIBentrySTDinterwordspacing

\bibitem{shamir1984identity}
A.~Shamir, ``Identity-based cryptosystems and signature schemes,'' in \emph{Workshop on the theory and application of cryptographic techniques}.\hskip 1em plus 0.5em minus 0.4em\relax Springer, 1984, pp. 47--53.

\bibitem{hu2010cryptanalysis}
X.~Hu, T.~Wang, and H.~Xu, ``Cryptanalysis and improvement of a hibe and hibs without random oracles,'' in \emph{2010 International Conference on Machine Vision and Human-machine Interface}.\hskip 1em plus 0.5em minus 0.4em\relax IEEE, 2010, pp. 389--392.

\bibitem{chow2004secureHIBS}
S.~S. Chow, L.~C. Hui, S.~M. Yiu, and K.~Chow, ``Secure hierarchical identity based signature and its application,'' in \emph{International Conference on Information and Communications Security}.\hskip 1em plus 0.5em minus 0.4em\relax Springer, 2004, pp. 480--494.

\bibitem{gentry2002hierarchical}
C.~Gentry and A.~Silverberg, ``Hierarchical id-based cryptography,'' in \emph{International conference on the theory and application of cryptology and information security}.\hskip 1em plus 0.5em minus 0.4em\relax Springer, 2002, pp. 548--566.

\bibitem{Dent2011}
\BIBentryALTinterwordspacing
A.~W. Dent, \emph{Certificateless Cryptography}.\hskip 1em plus 0.5em minus 0.4em\relax Boston, MA: Springer US, 2011, pp. 192--193. [Online]. Available: \url{https://doi.org/10.1007/978-1-4419-5906-5_314}
\BIBentrySTDinterwordspacing

\bibitem{HMACStandard}
\BIBentryALTinterwordspacing
{NIST}, ``{T}he {K}eyed-{H}ash {M}essage {A}uthentication code ({HMAC}),'' {U.S.} {D}epartment of {C}ommerce, Gaithersburg, MD, Standard FIPS PUB 198-1, 2008. [Online]. Available: \url{https://nvlpubs.nist.gov/nistpubs/FIPS/NIST.FIPS.198-1.pdf}
\BIBentrySTDinterwordspacing

\bibitem{HashStandard}
\BIBentryALTinterwordspacing
------, ``{R}ecommendation for {A}pplications {U}sing {A}pproved {H}ash {A}lgorithms,'' {U.S.} {D}epartment of {C}ommerce, Gaithersburg, MD, Standard NIST SP 800-107 Rev. 1, 2012. [Online]. Available: \url{https://nvlpubs.nist.gov/nistpubs/Legacy/SP/nistspecialpublication800-107r1.pdf}
\BIBentrySTDinterwordspacing

\bibitem{EP2661039B1}
\BIBentryALTinterwordspacing
S.~Gregory~T., ``Systems and methods for enhanced {ATC} overlay modulation,'' 20 Apr. 2016. [Online]. Available: \url{https://patents.google.com/patent/EP2661039B1}
\BIBentrySTDinterwordspacing

\bibitem{US20100079329A1}
\BIBentryALTinterwordspacing
------, ``Systems and methods for providing an advanced atc data link,'' 1 Apr. 2010. [Online]. Available: \url{https://patents.google.com/patent/US20100079329}
\BIBentrySTDinterwordspacing

\bibitem{Yeste-pskModulation-2015}
O.~Yeste-Ojeda and R.~Landry, ``\BIBforeignlanguage{English}{{ADS-B} {A}uthentication {C}ompliant with {M}ode-{S} {E}xtended {S}quitter {U}sing {PSK} {M}odulation},'' \emph{\BIBforeignlanguage{English}{2015 IEEE 18th International Conference on Intelligent Transportation Systems (ITSC). Proceedings}}, pp. 1773 -- 8, 2015.

\bibitem{Zambrano-PSK-2018}
A.-Q. Nguyen, A.~Amrhar, J.~Zambrano, G.~Brown, J.~Landry, R., and O.~Yeste, ``\BIBforeignlanguage{English}{Application of phase modulation enabling secure automatic dependent surveillance-broadcast},'' \emph{\BIBforeignlanguage{English}{Journal of Air Transportation}}, vol.~26, no.~4, pp. 157 -- 70, 2018.

\bibitem{Leonardi-phaseMod-2020}
M.~Leonardi and M.~Maisano, ``\BIBforeignlanguage{English}{Backward compatible physical layer protocol evolution for {ADS-B} message authentication},'' \emph{\BIBforeignlanguage{English}{IEEE Aerospace and Electronic Systems Magazine}}, vol.~35, no.~5, pp. 16 -- 26, 2020.

\bibitem{Arbuckle2021}
\BIBentryALTinterwordspacing
A.~Doug, ``Future {ADS-B} applications,'' ICAO, Technical Report Technical On-Line Workshop for the NAM/CAR Regions (ADS-B/OUT/W), 2021. [Online]. Available: \url{https://www.icao.int/NACC/Documents/Meetings/2021/ADSB/P05-FutureADS-B-ENG.pdf}
\BIBentrySTDinterwordspacing

\bibitem{SHA2}
\BIBentryALTinterwordspacing
{NIST}, ``{S}ecure {H}ash {S}tandard {(SHS)},'' {U.S.} {D}epartment of {C}ommerce, Gaithersburg, MD, Standard FIPS PUB 180-4, 2015. [Online]. Available: \url{https://nvlpubs.nist.gov/nistpubs/FIPS/NIST.FIPS.180-4.pdf}
\BIBentrySTDinterwordspacing

\bibitem{SHA3}
\BIBentryALTinterwordspacing
------, ``{SHA-3} {S}tandard: {P}ermutation-{B}ased {H}ash and {E}xtendable-{O}utput {F}unctions,'' {U.S.} {D}epartment of {C}ommerce, Gaithersburg, MD, Standard FIPS PUB 202, 2015. [Online]. Available: \url{https://nvlpubs.nist.gov/nistpubs/FIPS/NIST.FIPS.202.pdf}
\BIBentrySTDinterwordspacing

\bibitem{sciancalepore2019sos}
S.~Sciancalepore and R.~Di~Pietro, ``Sos: Standard-compliant and packet loss tolerant security framework for ads-b communications,'' \emph{IEEE Transactions on Dependable and Secure Computing}, vol.~18, no.~4, pp. 1681--1698, 2019.

\bibitem{encoder.py}
\BIBentryALTinterwordspacing
{Linar Yusupov}, ``{ADSB}\_{E}ncoder.py,'' 2017. [Online]. Available: \url{https://github.com/lyusupov/ADSB-Out}
\BIBentrySTDinterwordspacing

\bibitem{sun1090mhz}
J.~Sun, \emph{The 1090 Megahertz Riddle: A Guide to Decoding Mode S and ADS-B Signals}, 2nd~ed.\hskip 1em plus 0.5em minus 0.4em\relax TU Delft OPEN Publishing, 2021.

\bibitem{matlab}
\BIBentryALTinterwordspacing
C.~Moler, ``{MATLAB version 9.11.0.1873467 (R2021b) Update 3},'' 2021. [Online]. Available: \url{https://www.mathworks.com/products/matlab.html}
\BIBentrySTDinterwordspacing

\bibitem{ITU}
\BIBentryALTinterwordspacing
ITU-R, ``Spectrum occupancy measurements and evaluation,'' ITU, Technical report Report SM.2256-1(06/2016), 2016. [Online]. Available: \url{https://www.itu.int/pub/R-REP-SM.2256-1-2016}
\BIBentrySTDinterwordspacing

\bibitem{sun2020analyzing}
J.~Sun and J.~M. Hoekstra, ``Analyzing {A}ircraft {S}urveillance {S}ignal {Q}uality at the 1090 {M}egahertz {R}adio {F}requency,'' in \emph{Proceedings of the 9th International Conference for Research in Air Transportation}, 2020.

\bibitem{dataset1}
\BIBentryALTinterwordspacing
Open{S}ky, ``Open{S}ky {R}aw {D}ata,'' 2018. [Online]. Available: \url{https://opensky-network.org/datasets/raw/protected}
\BIBentrySTDinterwordspacing

\bibitem{Schaefer14a}
M.~Schäfer, M.~Strohmeier, V.~Lenders, I.~Martinovic, and M.~Wilhelm, ``Bringing {U}p {O}pensky: A {L}arge-scale {ADS-B} {S}ensor {N}etwork for {R}esearch,'' in \emph{Proceedings of the 13th International Symposium on Information Processing in Sensor Networks}.\hskip 1em plus 0.5em minus 0.4em\relax IPSN, 2014, pp. 83--94.

\bibitem{Kistan2017}
T.~Kistan, A.~Gardi, R.~Sabatini, S.~Ramasamy, and E.~Batuwangala, ``An evolutionary outlook of air traffic flow management techniques,'' \emph{Progress in Aerospace Sciences}, vol.~88, pp. 15--42, 2017.

\bibitem{sciancalepore2019reliability}
S.~Sciancalepore, S.~Alhazbi, and R.~Di~Pietro, ``Reliability of ads-b communications: Novel insights based on an experimental assessment,'' in \emph{Proceedings of the 34th ACM/SIGAPP Symposium on Applied Computing}, 2019, pp. 2414--2421.

\bibitem{TCAS2}
\BIBentryALTinterwordspacing
{FAA}, \emph{\BIBforeignlanguage{English}{{I}ntroduction to {TCAS II} {Version} 7.1}}, 2011. [Online]. Available: \url{https://www.faa.gov/documentLibrary/media/Advisory_Circular/TCAS%20II%20V7.1%20Intro%20booklet.pdf}
\BIBentrySTDinterwordspacing

\bibitem{ICAOPKD}
\BIBentryALTinterwordspacing
{ICAO PKD}, ``{ICAO} {P}ublic {K}ey {D}irectory {ICAO} {PKD} {W}hite {P}aper – {S}ystem {S}pecification for participants,'' ICAO, Montréal, QC, White paper, 2020. [Online]. Available: \url{https://www.icao.int/Security/FAL/PKD/Documents/PKDTechnicalDocuments/ICAO%20PKD%20White%20Paper_2020-07.pdf}
\BIBentrySTDinterwordspacing

\bibitem{ICAOPKDparticipants}
\BIBentryALTinterwordspacing
{ICAO Security and Facilitation}, ``{ICAO} {PKD} {P}articipants,'' 2022. [Online]. Available: \url{https://www.icao.int/Security/FAL/PKD/Pages/ICAO-PKDParticipants.aspx}
\BIBentrySTDinterwordspacing

\end{thebibliography}

\end{document}